\documentclass[pra,superscriptaddress,showpacs,twocolumn,nofootinbib,floatfix,amsmath,amssymb]{revtex4-1}
\usepackage{mathtools}
\usepackage{color}
\usepackage{cancel}
\usepackage{braket}
\usepackage{bm}
\usepackage{graphicx}
\usepackage{hyperref}

\def\mat#1{\bm{\mathrm{#1}}}
\def\op#1{\hat{#1}}

\newcommand{\ii}{\mathrm{i}}

\makeatletter
\@ifundefined{textcolor}{}
{%
 \definecolor{BLACK}{gray}{0}
 \definecolor{WHITE}{gray}{1}
 \definecolor{RED}{rgb}{1,0,0}
 \definecolor{GREEN}{rgb}{0,1,0}
 \definecolor{BLUE}{rgb}{0,0,1}
 \definecolor{CYAN}{cmyk}{1,0,0,0}
 \definecolor{MAGENTA}{cmyk}{0,1,0,0}
 \definecolor{YELLOW}{cmyk}{.2,.4,1,0}
 \definecolor{orange}{rgb}{1,0.5,0}
 }
\makeatother

\begin{document}

\title{What does it mean for half of an empty cavity to be full?}
\author{Eric G. Brown}
\affiliation{Department of Physics \& Astronomy, University of Waterloo, Waterloo, ON, N2L 3G1, Canada}
\author{Marco del Rey}
\author{Hans Westman}
\author{Juan Le\'{o}n}
\affiliation{Instituto de F\'{i}sica Fundamental, CSIC, Serrano 113-B, 28006 Madrid, Spain}
\author{Andrzej Dragan}
\affiliation{Institute of Theoretical Physics, University of Warsaw, Pasteura 5, 02-093 Warsaw, Poland}

\begin{abstract}
It is well known that the vacuum state of a quantum field is spatially entangled. This is true both in free and confined spaces, for example in an optical cavity. The obvious consequence of this, however, is surprising and intuitively challenging; namely, that in a mathematical sense half of an empty cavity is not empty. Formally this is clear, but what does this physically mean in terms of, say, measurements that can actually be made? In this paper we utilize the tools of Gaussian quantum mechanics to easily characterize the reduced state of a subregion in a cavity and expose the spatial profile of its entanglement with the opposite region. We then go on to discuss a thought experiment in which a mirror is introduced between the regions. In so doing  we expose a simple and physically concrete answer to the above question: the vacuum excitations resulting from entanglement are mathematically equivalent to the real excitations generated by suddenly introducing a mirror. Performing such an experiment in the laboratory may be an excellent method of verifying vacuum entanglement, and we conclude by discussing different possibilities of achieving this aim.
\end{abstract}
\maketitle
 
\section{Introduction}

The elementary excitations of quantum field theory
are countable, a crucial feature for making it able to deal
with the physics of elementary particles. These are described
by operators which carry information about the
energy momentum that these excitations take or give to
the field. The ground state of the field, from which no
quanta can be removed, thus becomes the vacuum, whose excitations describe the states with one or
more particles with well defined momenta. This Fock
construction provides simple global operators for the field
as a whole. In particular, the total number of particles
carried by any specified configuration is easy to address.
However, the construction lacks a means to inquire
into the local properties of the field. For instance a question
as simple as ``how are these particles distributed in space?" or
more simply, ``where is this particle?" are difficult to address
if not by indirect means. The root of the problem is
that creation and annihilation operators do not belong to
the algebra of local operators of the quantum field. This
can be understood in physical terms as a consequence
of the fact that they describe excitations endowed with
sharp momentum and hence expected in principle to be
unlocalized in space.
The conclusion is that, powerful as it is, the standard Fock space techniques provide only a feeble scaffolding for digging into issues pertaining
to the localization of quanta. 

These issues are also deeply connected to the well-known fact that the vacuum state of a quantum field displays quantum entanglement between space-like separated regions \cite{Summers87,Srednicki93}. Much work has been performed, using a variety of mathematical approaches and models, to understand and characterize the properties of this entanglement \cite{Casini09}. In addition, it has been proposed that this entanglement may be ``harvested" (i.e. swapped) to  an auxiliary quantum system without the need for causal interaction \cite{Reznik05,Braun05,Leon09,Steeg09,Brown13-3}, which may then be used for quantum informational procedures. The existence of spatial entanglement is similarly present in condensed matter and lattice systems \cite{Cramer06,Amico08}, being a generic property of extended systems with local interactions, of which a quantum field can simply be viewed as a continuum limit. While in such systems experimental proposals have been put forth for the verification of vacuum entanglement (e.g. a pair of trapped ions) \cite{Retzker05}, to the authors' knowledge no feasible concrete proposal has yet been suggested for its verification in a true, relativistic, quantum field (e.g. the photon field).  

One immediate consequence of vacuum entanglement is that, due to the vacuum being a pure state, the reduced state over any local region in space must necessarily be mixed and thus excited. Relativistic quantum phenomena involving the observer-dependence of particle number, such as the Unruh and Hawking effects, are often attributed to this \cite{BirrelDavies84}. Moreover, vacuum entanglement occurs also in enclosed systems, such as an optical cavity or a superconducting circuit. This introduces a conceptually challenging fact: at least formally, \emph{half of an empty box is not empty}. This is a mathematical result which alone gives us little intuition towards actual physical consequences. Under what operational conditions does this phenomenon present itself; what physically sensible measurements (in general) can be made to give this mathematical fact experimental significance? 

If an experimentalist has such an empty cavity then what can they do to detect photons in, say, the left half of the cavity (that supposedly contains many)? The answer, as will be explained, is to very quickly introduce a physical boundary (in this case a mirror) between the two sides of the cavity, thus blocking off any influence from the right-hand side while the experimentalist measures the left-hand side. Of course, quickly introducing a boundary (i.e. quickly modifying the Hamiltonian) produces particles similar to what occurs in the dynamical Casimir effect (DCE) \cite{Moore70}, which has recently been experimentally observed \cite{Wilson11}. The key observation of this paper, however, is that these real excitations, created by slamming down the mirror, are mathematically equivalent to the vacuum excitations of the reduced state that we attribute to entanglement. This is what it operationally means for half of an empty box to be non-empty. In addition to giving a satisfying interpretation to the problem of local particle content, we will discuss how this realization provides a simple experimental setup that can be used to measure, and perhaps even utilize, local vacuum excitations. 

In this paper we consider both massive and massless scalar fields in a one-dimensional cavity with Dirichlet boundary conditions (i.e. mirrors on the ends). We perform several tasks. To begin, we discuss the difficulties that appear when we intend to measure vacuum excitations and the different alternative scenarios that could allow us to circumvent them. We will present a recently introduced formalism of local quantization \cite{Vazquez14} that allows us to characterize the reduced state of a sub-region in the cavity and study its local properties formally. At that point we will consider what occurs if a mirror is very quickly introduced in the middle of the cavity. 

We will utilize Gaussian quantum mechanics \cite{Adesso07} in order to easily compute and characterize the reduced states of sub-cavity regions and the correlations between them, explaining how this equivalently describes the physics of slamming mirror(s) into the cavity. We will discuss and analyse the spatial structure of entanglement between regions, similar to what has been done in \cite{Botero04} for lattice systems. We will furthermore discuss the time-evolution of the system after slamming a mirror and observe what one would expect: a burst of particles propagating away from it. These excitations, however, are mathematically one and the same with those previously attributed to vacuum entanglement in the local analysis (the only difference is that now they evolve according to a different Hamiltonian). Equivalently, the real excitations produced in the left and right-hand sides are quantum entangled. We also consider the case in which two mirrors are simultaneously introduced, some distance apart. In this case the particles produced in the left and right-hand sides (but ignoring the middle section) can similarly be entangled, despite no common mirror between them. This is possible because, as follows from the main point of our paper, the entanglement is simply that which was already present in vacuum prior to the introduction of the mirror.

Lastly, we discuss the experimental feasability of using this scenario to verify vacuum entanglement using current technologies. We note that introducing a mirror in fact represents an very efficient means of harvesting the vacuum entanglement, since afterwards you have two new cavities that contain \emph{all} of the entanglement (up to a UV cut-off determined by how fast the mirror is introduced). This entanglement could then be a resource for quantum computational tasks. This method of harvesting could potentially be much more promising than the usual proposed method of locally interacting a pair of other quantum systems (e.g. artificial atoms) with the field  \cite{ReznikRetzker05,Steeg09, EduEric13}, since this is severely limited by the interaction strength.

Throughout this paper we will work in natural units such that $\hbar = c = k_B = 1$.

\section{How does one measure the vacuum excitations in a subregion?}   \label{how}

To begin, we need to ask ourselves in a general sense what one must do in order to measure localized vacuum fluctuations. What operational procedures can be implemented to do this? Mathematically these fluctuations can arise when tracing out a spatial region of a vacuum field. That is, because there is entanglement between spatial regions, the reduced state of such a region must necessarily be mixed (and therefore excited). This thus motivates the idea that at least one possible way of measuring these excitations is to isolate oneself to only the subregion of interest. But this means more than simply staying at a fixed location. As we will show later in more detail, a stationary detector interacting with a vacuum field only at a given point or region will still register zero particle detection if it is allowed to measure for a long enough time. Rather, isolating oneself to only a subregion means losing causal contact with the outside; information cannot be allowed to reach our observer from outside the region of interest. Uniform acceleration, for example, is one way of achieving this \cite{Unruh, BirrelDavies84}. Another way is for one to turn their detection device on for only a short time $\Delta t$; doing this ensures that the detector is causally isolated from any part of space more than a distance $c\Delta t$ away from it. Indeed switching one's detector on fast enough does cause spurious detection events \cite{Brown13, delRey12} , however it is questionable if this can be fully attributed to vacuum excitations (i.e. to entanglement) inside a cavity since formally the probability of detection limits to zero only as $\Delta t \rightarrow \infty$, which is clearly larger than the cavity length.

Are there any other ways to isolate oneself from outside influence? Indeed, another option that gets the job done is simply to erect a physical boundary. In the cavity scenario this corresponds to placing a perfect mirror at the bipartition boundary. Certainly once such a mirror is installed then an observer in the left-side of the cavity will receive no information from the right-side. Would such an observer then be free to measure local vacuum excitations?  How could it be that such a setup suddenly allows the observer to measure what they could not have beforehand? Furthermore, one should be concerned about the fact that quickly placing a mirror in the middle of the cavity is expected to create real particle excitations, similar to what occurs in the dynamical Casimir effect (DCE) \cite{Moore70,Wilson11}. That is, by rapidly changing (in this case, introducing) a boundary condition we are rapidly modifying the Hamiltonian of the system. This will create real excitations in the field that will propagate away from the mirror upon being introduced, and an observer located on one side of this mirror will detect these excitations. Will these particles interfere with the observer's ability of detect the local vacuum excitations that are associated with entanglement between regions?

The answer, as we will elaborate, is that a detection of the mirror-created particles \emph{is} exactly a detection of the local entanglement excitations.

\begin{figure}[t]
	\centering
    \includegraphics[width=0.48\textwidth]{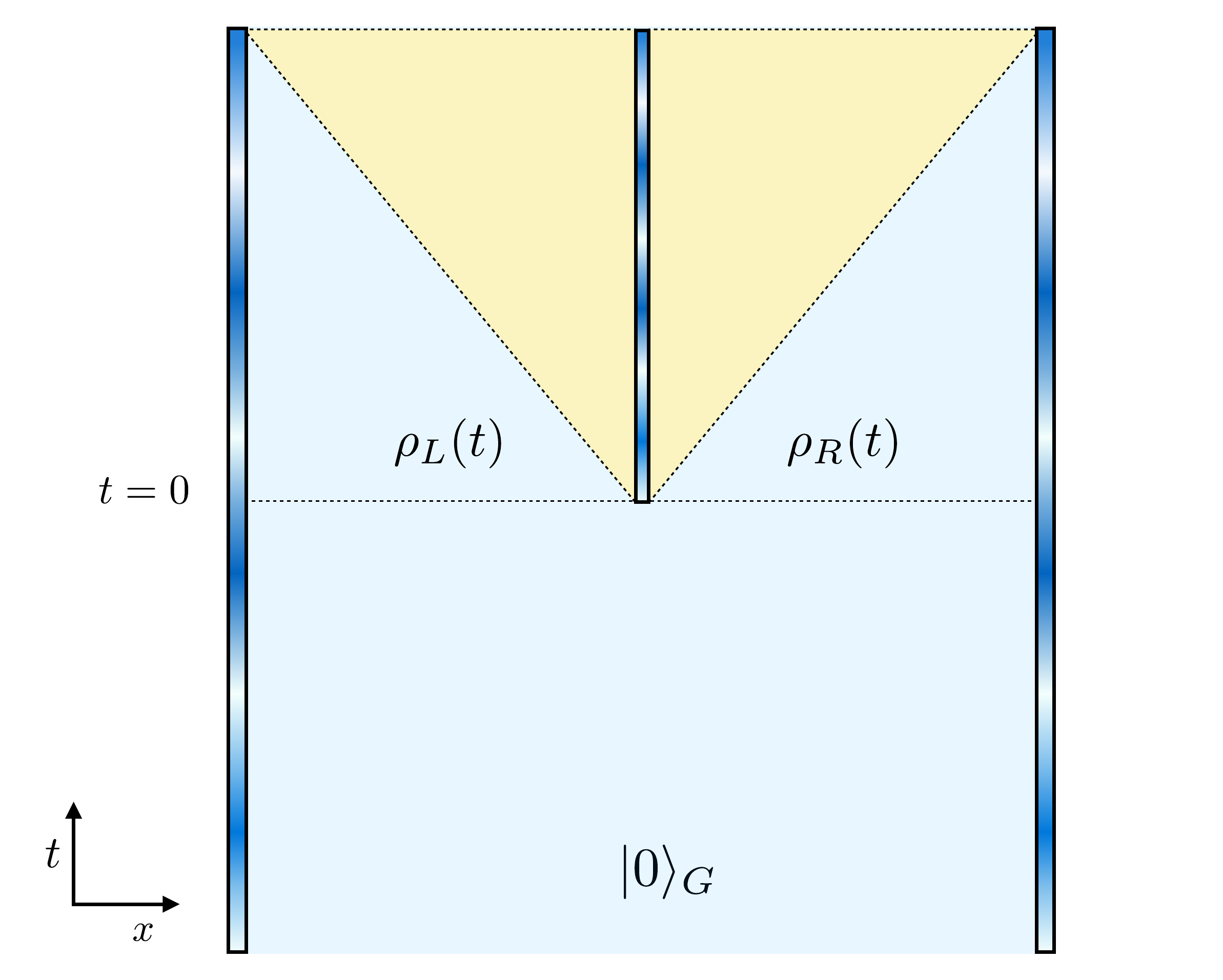}
	\caption{Sketch of the one dimensional cavity setting. We start ($t<0$) considering a cavity in its vacuum state $|0\rangle_G$. At some instant ($t=0$) we slam a mirror  separating the cavity into two regions. As explained in the text, the normal modes in these separated subcavities correspond to the localized modes of the cavity without mirror, which we will show suffice to analize states, correlations and particle production after slamming the mirror. The horizontal line corresponds to $t=0$, the diagonal lines represent the light cone starting at the slamming event.}
        \label{spacetimediagram}
\end{figure}

\section{Formulation and setting}



The first purpose of this section is to present a mathematical framework for the computation and analysis of global cavity states using a local formalism. We will start describing the quantum states in the cavity by introducing a bipartition of it into two subregions, precisely those in which the cavity will eventually be separated by the introduction of a slamming mirror at some instant of time. Later on, in the second part of this section, we will use this formalism to study the physical scenario where a mirror is abruptly introduced in the middle of the cavity.

\subsection{Local mathematical analysis: local vs. global modes}  \label{LVG}

Here we will briefly introduce the field-theoretic formalism required for our analysis \cite{BirrelDavies84,Vazquez14}. 

The aim is to spell out what can be ascertained about the physics of a non-localized state spanning the whole cavity, as is the case of the quantum vacuum and generic cavity states, by using localizing mathematical tools. We do not yet introduce a mirror in the middle of the cavity. We will postpone this to the next subsection, once the present goal is achieved.

Let us consider a quantum scalar field $\op \phi(x,t)$ of mass $\mu$ within a cavity of length $R$, such that $x \in [0, R]$. Specifically, we will consider a cavity with Dirichlet (i.e. mirror) boundary conditions, as would be the case in a physical optical cavity, for example. The field is thus constrained to satisfy $\op \phi(0,t)=\op \phi(R,t)=0$.

Our field can be expanded in the standard form:
\begin{align}
	\op \phi(x,t)=\sum_{m=1}^\infty (f_m(x,t) \op b_m + f^*_m(x,t) \op b_m^\dagger).
\end{align}
Here, the set of chosen mode functions $\{f_n\}$ are required to satisfy the Klein-Gordon equation $(\Box+\mu^2)f(x,t)=0$ as well as the correct boundary conditions. In addition, they must form a complete and orthonormal set with respect to the Klein-Gordon inner product \cite{BirrelDavies84}. Aside from these constraints the choice is arbitrary. Making such a choice is equivalent to a choice of Fock basis, for which the operators $\{\op b_n, \op b_n^\dagger\}$ are the corresponding ladder operators.

The standard choice for a Dirichlet cavity, which we will refer to as the \emph{global modes} $U_n$, are given by
\begin{align}   \label{globalmodes}
	U_n(x,t)=\frac{1}{\sqrt{R \Omega_n}}\sin \left(\frac{\pi n x}{R}\right)e^{-\ii \Omega_n t} = \mathcal U_n (x) e^{-\ii \Omega_n t} ,
\end{align}
where $\Omega_n^2=\frac{\pi^2 n^2}{R^2}+\mu^2$ is the frequency of mode $n$. This choice is convenient because the corresponding Fock states are energy eigenstates of the free-field regularised Hamiltonian (which we will also call the \emph{global Hamiltonian})
\begin{align}
	\op H_G=\sum_{n=1}^\infty \Omega_n \op A_n^\dagger \op A_n,
\end{align}
where here $\{\op A_n,\op A_n^\dagger\}$ are the ladder operators corresponding to the global modes. A state of principal importance for us is the \emph{global vacuum state} $\ket{0_G}$, defined to satisfy $\op A_n \ket{0_G}=0$ for all $n$. This is the state of lowest energy with respect to $\op H_G$, and is said to be the state of zero particles, because no A quanta can be removed from it, i.e.  a cavity in this state is empty (although not from a local point of view as we discuss later). 

While the field decomposition into the global modes is often the most convenient and physically relevant choice, we can also consider a decomposition into a mode basis better suited to study the local physics of a subregion inside the cavity. Say that we decompose our cavity into two regions, one that runs within $x \in [0,r]$ (the left side) and the other within $x \in [r,R]$ (the right side). The lengths of these two sides are thus $r$ and $\bar r \equiv R-r$, and we can define a new set of modes $\{u_m(x,t)\}$ and $\{\bar{u}_m(x,t)\}$ for the left and right sides, respectively. The obvious way of doing this is to define these modes to have support at a certain time $t=0$ only over their corresponding subregions. As pointed out in \cite{Vazquez14}, however, one must be careful that the new basis modes still satisfy the correct boundary conditions of the cavity (and in particular, not extra ones). This requirement immediately implies that if, say the set $\{u_m\}$ are supported only in the left region at $t=0$, then their support will necessarily exceed this region for later times ($u_m$ satisfies the wave equation and we have not placed an extra boundary condition between the two regions). This does not turn out to be a hindrance in exploring local physics, however. 

Since the global vacuum $\ket{0_G}$ is a stationary state it does not matter at what time we examine its properties; we will choose time $t=0$. The solution is then to simply define our local modes to be appropriately compactly supported at this instant. To this end, we will define our local modes $u_m(x,t)$ to have initial conditions
\begin{align}          \label{localmodes}
	&u_m(x,0) =\frac{\theta(r-x)}{\sqrt{r\omega_m}}\sin \left(\frac{\pi m x}{r}\right), \nonumber \\
	 &\dot u_m(x,0)=-\ii \omega_m u_m(x,0), \nonumber  \\
	&\bar u_m(x,0) =\frac{\theta(x-r)}{\sqrt{\bar r \bar \omega_m}}\sin \left(\frac{\pi m (x-r)}{\bar r} \right),  \nonumber \\
	&\dot{\bar u}_m(x,0)=-\ii \bar \omega_m \bar u_m(x,0),
\end{align}
where $\omega_m^2=\frac{\pi^2 m^2}{r^2}+\mu^2$ and $\bar \omega_m^2=\frac{\pi^2 m^2}{\bar r^2}+\mu^2$. 
Given the above initial conditions, the local modes will evolve throughout the cavity according to the Klein-Gordon equation. These modes satisfy the proper boundary conditions and constitute a complete and orthonormal basis for the whole cavity \cite{Vazquez14}, and thus form a proper expansion of the field. For our purposes in this section, however, we need only consider the instant $t=0$ at which they are localized to their respective sides of the cavity. Examining the global vacuum in this basis, at this instant, allows us to fully characterize the reduced state of the subregions and the quantum correlations between them. The decomposition in terms of local modes is depicted in Fig. \ref{slamming}.

Let us denote the local ladder operators associated with the above modes as $\{\op a_m, \op a_m^\dagger\}$ for the left side, and $\{\op{\bar a}_m, \op{\bar a}_m^\dagger\}$ for the right.

Solutions sets to the Klein-Gordon equation are related by a linear \emph{Bogoliubov transformation} \cite{Bogos,BirrelDavies84}. This means that our local modes are related to the global modes via some transformation of the form
\begin{align}
	u_m(x,t)&=\sum_{n=1}^\infty (\alpha_{mn}U_n(x,t)+\beta_{mn}U_n^*(x,t)), \nonumber \\ 
	\bar u_m(x,t)&=\sum_{n=1}^\infty (\bar \alpha_{mn}U_n(x,t)+\bar \beta_{mn}U_n^*(x,t)).	
	\label{eq:bogoliubovtrans}
\end{align}
Equivalently, in terms of the annihilation operators (from which the creation operators are trivially obtained) we have
\begin{align}  \label{ladderTrans}
	\op a_m&=\sum_{n=1}^\infty (\alpha^*_{mn} \op A_n - \beta^*_{mn}\op A_n^\dagger),  \nonumber \\
	\op{\bar a}_m&=\sum_{n=1}^\infty (\bar \alpha^*_{mn} \op A_n - \bar \beta^*_{mn}\op A_n^\dagger).
\end{align}
The \emph{Bogoliubov coefficients}, which are time-independent, are computed via the Klein-Gordon inner products between local and global modes. In our case, they evaluate to \cite{Vazquez14}

\begin{alignat}{2}\label{bogos}
 \alpha_{mn} &= (U_n | u_m)  &=  (\Omega_n+\omega_m) \mathcal V_{mn},  \\
 \beta_{mn} &= - (U^*_n|u_m) &= (\Omega_n-\omega_m) \mathcal V_{mn}, \\
 \bar{\alpha}_{mn} &= (U_n | \bar u_m)  &=  (\Omega_n+\omega_m) \bar{\mathcal V}_{mn},  \\
 \bar{\beta}_{mn} &= - (U^*_n| \bar u_m) &= (\Omega_n-\omega_m)\bar{\mathcal V}_{mn},
\end{alignat}

where

\begin{alignat}{2}
  \mathcal V_{mn}  &= \int_0^R \text{d}x\,\mathcal U_n(x) u_m(x,0)  \\ &= \frac{\frac{m\pi}{r}(-1)^m}{\sqrt{Rr\Omega_n\omega_m}(\Omega_n^2-\omega_m^2)} \sin\frac{n\pi r}{R}, \\
 \bar {\mathcal V}_{mn}  &= \int_0^R \text{d}x\, \mathcal U_n(x) \bar u_m(x,0)  \\&= \frac{-\frac{m\pi}{\bar r}}{\sqrt{R\bar r\Omega_n\bar\omega_m}(\Omega_n^2-\bar\omega_m^2)}\sin\frac{n\pi r}{R}.
\end{alignat}

\begin{figure}[t]
	\centering
    \includegraphics[width=0.48\textwidth]{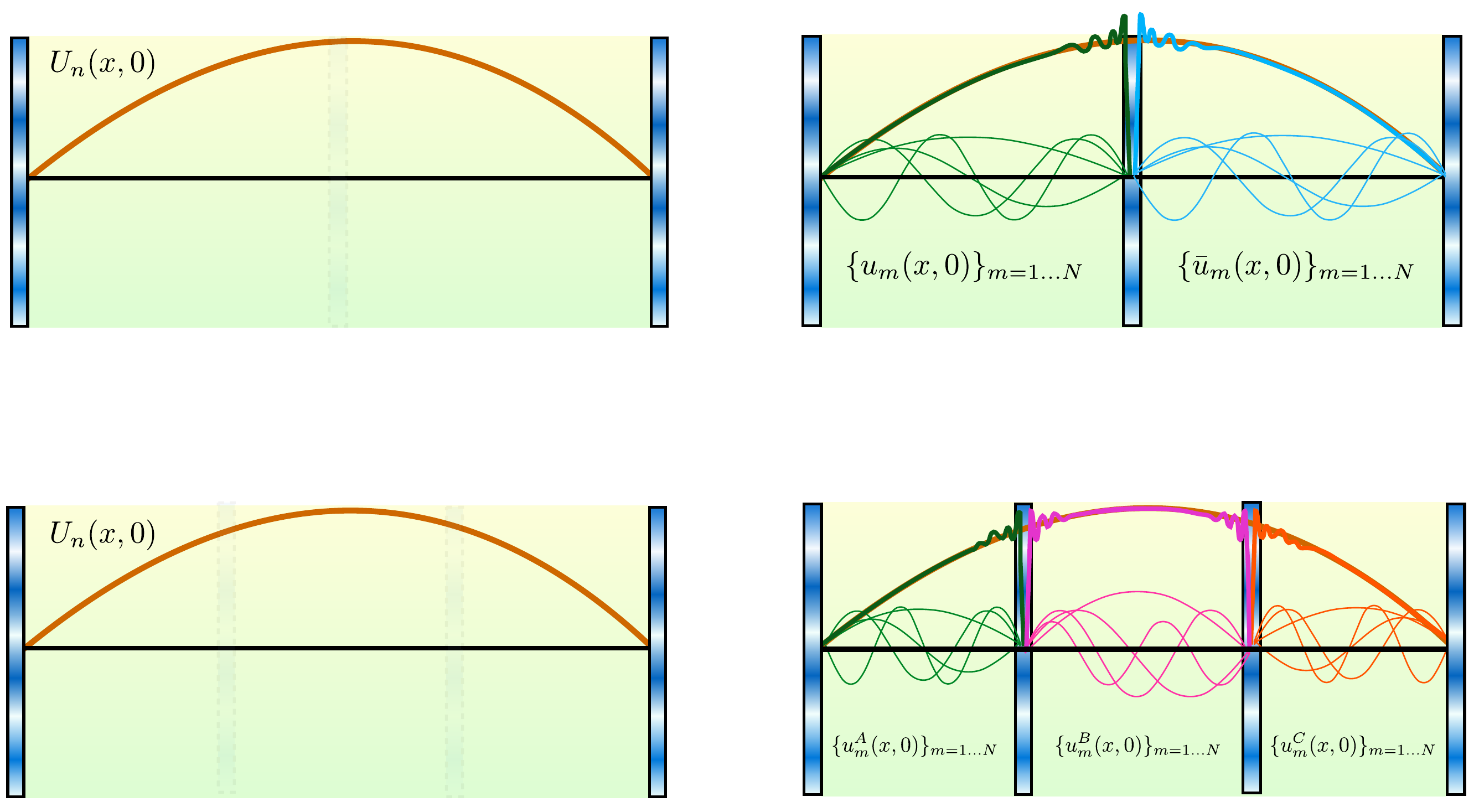}
	\caption{The cavity for the cases studied in the paper. The figures on the left correspond to the full cavity without mirrors, the light dotted vertical bars indicating the border of the regions chosen in sections IIIA and IIIC to study localization into two (top) or three (bottom) spatial regions. The right figures show the decomposition in terms of local modes at $t=0$ for both settings. }
        \label{slamming}
\end{figure}
�

The fact that the $\beta$ coefficients are non-vanishing implies that the global vacuum $\ket{0_G}$ is, in the local basis, an excited state in the sense that $\op a_m \ket{0_G} \neq 0$ and $\op{\bar a}_m \ket{0_G} \neq 0$, i.e. local quanta can be removed from it, so in this picture the vacuum cannot be considered to be empty. Indeed, the reduced state of, say, the left side of the cavity, is a mixed state. These local excitations, and the local mixedness, are associated with the entanglement present between the two sides of the cavity. 

Lastly, as with any Bogoliubov transformation, the above coefficients satisfy the necessary conditions \cite{BirrelDavies84}
\begin{align}  \label{Wron1}
	\sum_k (\alpha_{mk}\alpha_{nk}^*-\beta_{mk}\beta_{nk}^*)&=\delta_{mn},   \\   \label{Wron2}
	\sum_k (\alpha_{mk}\beta_{nk}-\beta_{mk}\alpha_{nk})&=0,
\end{align}
and similarly for the barred coefficients.

\subsection{Slamming down a mirror}   \label{slamming}
If we compute the vacuum expectation value of the local number operators $\op n_m=\op a_m^\dagger a_m$ and $\op{\bar n}_m=\op{\bar a}_m^\dagger \op{\bar a}_m$ we find that these are non-zero for the global vacuum state $|0_G\rangle$, indicating the presence of a bath of `local particles'. While this observation is mathematically correct we must nevertheless ask ourselves if any operational significance can be attached to this theoretical notion of ``local quanta". Can we somehow detect the presence of such local quanta in the lab?

Taking as inspiration the discussion in Sect. \ref{how}, we claim that a generic (but perhaps not exhaustive) method of achieving this is to informationally block the local region of interest from the rest of the system. In a cavity-field system this can be achieved by introducing a mirror, separating the cavity into two new smaller ones. Indeed, as we will discuss, such an operation \emph{does} allow the detection and characterization of local excitations. This is fundamentally due to the fact that we identify a ``real" (i.e. measurable) particle to be an elementary excitation of a \emph{stationary} field mode. By the act of introducing the mirror, what were nonstationary local modes of the full cavity translate into stationary modes of the new small cavity, thus facilitating the measurement of their excitations.  

One may be concerned with the unique identification of ``a real particle" with ``an elementary excitation of a stationary mode". In this work, however, we attempt to be operationally unambiguous and connect as closely as possible with the kinds of measurements that can actually be achieved in the laboratory, necessitating long measurement times as compared to the fundamental time-scale of the cavity. As a detection model let us consider an idealized, point-like, DeWitt monopole detector that remains at some specific location $x_0$. The following observation, however, is valid for \emph{any} choice of detector. Let the initial state of the system is taken to be the $|0_G\rangle\otimes |g\rangle$ where $|g\rangle$ is the ground state of the detector. We will present two cases:

First, \textit{without slamming a mirror}, we imagine adiabatically switching on the coupling between field and detector. The adiabatic theorem guarantees that if the system was originally in the ground state of the non-interacting theory, then the system at much later times will be found in the interacting ground state.\footnote{The adiabatic theorem requires a gap between the vacuum energy eigenvalue and other eigenvalues. This is guaranteed since we are dealing with a cavity with a naturally infrared cut-off defined by the size of the cavity $R$.} When we then adiabatically switch off the interaction the combined system will then be found in the non-interacting vacuum and thus will fail to detect the presence of local quanta. This then immediately shows that such a detector will not get excited. Such a procedure does not detect particles when the global system is in its ground state (thus allowing us to use the adiabatic theorem). It is for this reason that we relate the particle notion with the free stationary modes, which are the ones corresponding to the free eigenstates.

If, on the other hand, \textit{we slam down a mirror} and then follow the same adiabatic detection procedure within one of the sub-cavities then we \emph{will} detect the presence of particles. This is because the local modes, which are stationary after the mirror is introduced, are excited. Critical to the message of this paper is that the measurement statistics that will be obtained from this procedure are equivalent to the local virtual particle statistics (i.e. those corresponding to one half of the box) obtained from the transformation presented in Sect. \ref{LVG}, which simply describes the local physics of the cavity and does not assume the introduction of a mirror.

Concretely, we consider what happens when we instantaneously introduce a mirror at $x=r$ and $t=0$, i.e. we impose the Dirichlet boundary condition $\phi(r,t)=0$ for $t\geq 0$. Clearly the instantaneous assumption is not physically realistic, however this turns out not to be a hindrance in elucidating the most realistic, finite-time case. This will be further discussed in Sect. \ref{finitetime}. Given this scenario, it is clear that the set of local modes with initial conditions \eqref{localmodes}, which were non-stationary for $t<0$ prior to the introduction of the mirror, will have a different evolution than before, $u'_m(x,t)$ and $\bar u'_m(x,t)$, which for $t\geq0$ will correspond to {\em stationary} modes , i.e.
\begin{align}
u'_m(x,t)= \begin{dcases*}
        u_m(x,t) & when $t\leq 0$ \\
        u_m(x,0) e^{-i\omega_m t} & when $t>0$
        \end{dcases*}\\
\bar u'_m(x,t)= \begin{dcases*}
        \bar u_m(x,t) & when $t\leq 0$ \\
        \bar u_m(x,0) e^{-i\bar \omega_m t} & when $t>0$
        \end{dcases*}
\end{align}
Please note that after this section we will only need to consider the times $t \geq 0$, and thus will abuse notation by dropping the primes from the mode functions, meaning that for $t>0$ we will define $u_m(x,t)=u_m(x,0)e^{-\ii \omega_m t}$. 

Furthermore and analogously, the corresponding global modes $U'_m(x,t)$ would only be stationary modes for $t<0$; for $t\geq0$ these modes would be non-stationary.

Equivalently, the sudden introduction of the mirror translates mathematically into a time-dependent Hamiltonian, i.e. we have
\begin{eqnarray}
\op H=\left\{ \begin{array}{cc}\sum_m \Omega_m \op A_m^\dagger   \op A_m  & t<0\\
\sum_m \omega_m \op{a}_m^\dagger  \op{a}_m+\bar\omega_m \op{\bar a}_m^\dagger  \op{\bar a}_m  & t\geq0 \end{array}\right. .
\end{eqnarray}
Physically, the time dependence of the Hamiltonian will cause particle creation similar to the Dynamical Casimir Effect \cite{Moore70,Wilson11}. 

To determine exactly the amount of particle creation we need to calculate the Bogoliubov coefficients between the modes $U'_m$ and $u'_m$ and $\bar u'_m$. These are nothing but the Klein-Gordon inner products $\alpha_{mn}=(u'_m|U'_n)$, $\beta_{mn}=(u'_m|U_n'^*)$, $\bar \alpha_{mn}=(\bar u'_m|U'_n)$, and $\bar \beta_{mn}=(\bar u'_m|U_n'^*)$, which we can conveniently evaluate at time $t=0$. Importantly, due to the specific choice of initial data at $t=0$ these Bogoliubov coefficients necessarily coincides with those of the previous section, i.e. Eq. \eqref{bogos}. This means that the particle content generated by the mirror is exactly equivalent to the local particle content ($a^\dagger_m a_m, \bar a^\dagger_m \bar a_m$) before the mirror is introduced, i.e. the particle content that is associated with entanglement between the two sides of the cavity. Thus, although the sudden introduction of a mirror is usually understood as causing particle creation, it is at the same time an operation that does not change the \emph{local} particle number of the state. The difference now being that these particle contents are associated with stationary modes, meaning that they can be measured using standard techniques of quantum optics.

Moreover, it is not just the particle content, but the state in general that does not change. That is, all particle statistics and correlations (including entanglement) are unchanged by the action of slamming the mirror. Slamming the mirror does of course change the time evolution of the system. For $t<0$ the system is time-independent, the global vacuum state being stationary with respect to the global Hamiltonian, whereas for $t>0$ the change of Hamiltonian will cause time evolution (e.g. particles propagating away from the mirror). The key observation, however, is that this difference in evolution is fully encompassed by the difference in spatial evolution of the mode functions themselves and not by any changes in particle content or correlations between them. 

%
%

\subsection{Three spatial regions}  \label{three1}

Before continuing we would also like to describe the case in which the cavity is split into three spatial regions, rather than only two. This will prove useful later when we discuss the operational implications of slamming down mirrors and the possible experimental verification of vacuum entanglement. Note that the extension to any number of regions follows analogously.

Let us proceed by considering a division of our cavity into three sections $\Delta_A = [A_0, B_0], \Delta_B= [B_0, C_0],$ and $\Delta_C=[C_0,R]$ with  sizes $A, B$ and $C$ respectively. Let us define: 
  \begin{displaymath}
  \Pi_{Z}(x)  = \left\{
     \begin{array}{lr}
       1 & : x \in \Delta_Z\\
       0 & : x \notin \Delta_Z
     \end{array}
   \right.,
\end{displaymath} 
and set  
\begin{align}
A_0 = 0,\quad B_0=A,\quad C_0=A+B,\quad R=A+B+C.
\end{align}

We can build the local modes for these three regions $u_l^{Z}(x,t)$ by demanding that:  
 \begin{align}
u_l^{Z} (x,0)&=\frac{\Pi_{Z}(x)}{\sqrt{Z\omega^Z_l}}\sin\frac{l \pi (x-Z_0)}{Z},\\
 \dot u^Z_l(x,0)&=-i\omega^Z_l u^Z_l(x,0).
\end{align}
 
 With: 
 \[\omega^Z_l =\sqrt{\left(\frac{m\pi}{Z} \right)^2 + \mu^2},\qquad Z = A,B,C.\]
 
The new Bogoliubov transformation, analogous to Eq. (\ref{eq:bogoliubovtrans}), is: 
\begin{align}
u^Z_m = \sum_n \alpha^Z_{mn} U_n +  \beta^Z_{mn} U^*_n, \qquad Z = A,B,C,
\end{align}
where 
\begin{alignat}{3}
& \alpha^Z_{mn} &&= (U_n | u_m^Z)  &&=  (\Omega_n+\omega^Z_m) \mathcal V_{mn}^Z,  \\
& \beta^Z_{mn} &&= - (U^*_n|u_m^Z) &&= (\Omega_n-\omega^Z_m) \mathcal V_{mn}^Z,
\end{alignat}
 and
\begin{align}
  \mathcal{V}^A_{mn}  &= \int_0^A \text{d}x\ \mathcal U_n(x) u^A_m(x,0) = \nonumber\\&\frac{\frac{m\pi}{A}(-1)^m}{\sqrt{RA\Omega_n\omega^A_m}({\Omega_n^2-\omega_m^{A^2}})}\sin\frac{n\pi A}{R}, \\
  \mathcal{V}^B_{mn}  &= \int_A^{A+B} \text{d}x \ \mathcal  U_n(x) u^B_m(x,0)   =\nonumber\\&\frac{\frac{m\pi}{B} \left[ (-1)^m \sin\left(\frac{n\pi(A+B)}{R}\right)  -\sin \left(\frac{n\pi A}{R}\right)\right] }{\sqrt{RB\Omega_n\omega^B_m} (\Omega_n^2-\omega_m^{B^2}) },  \\
  \mathcal{V}^C_{mn}  &= \int_{A+B}^R \text{d}x \ \mathcal  U_n(x) u^C_m(x,0)  =\nonumber\\&\frac{-\frac{m\pi}{C}}{\sqrt{RC\Omega_n\omega^C_m}(\Omega_n^2-\omega_m^{C^2})}\sin\frac{n\pi (A+B)}{R}.
\end{align}

Transforming to this mode basis allows us to describe the local physics of, and the correlations between, these three regions. Similar to the scenario discussed in Sect. \ref{slamming}, the mode basis described here can be used to describe the process of slamming down two mirrors simultaneously, thereby splitting the cavity into three regions. Exactly the same physics applies in this case, and we will thus not reiterate the material of Sect. \ref{slamming}.

\section{Computing the state}    \label{Computestate}

In this section we will focus on obtaining a local description of the global vacuum state \footnote{It must be pointed out that the mathematical toolbox presented here allow us to work with any Gaussian state, not just the global vacuum. We could, for example, start with with a global thermal state.}.  This includes the evaluation of the reduced field state of a subregion of a cavity, and a description of the of vacuum entanglement between regions of the cavity. We rely on the formalism of Gaussian quantum mechanics \cite{Adesso07} for our exposition. The unfamiliar reader is encouraged to read Appendix. \ref{Gauss}, which outlines the concepts of Gaussian quantum mechanics that are necessary to understand the main text. A key point to keep in mind, as discussed in the previous section, is that the Bogoliubov transformation (and thus the resulting state) is the same whether we consider this to be with or without the introduction of the mirror. As discussed further in Sect. \ref{timeevolution}, the covariance matrix that we compute (i.e. the state) equally well describes both cases.


\subsection{The state of two regions}

We will start by computing the form of the global vacuum upon transforming to the local-mode basis, in the case that we split the cavity into two regions. Let us define the canonically conjugate quadrature operators for the field modes, both global and local. Letting $\{\op A_n,\op A_n^\dagger\}$ be the ladder operators for the global modes, we define the corresponding quadrature operators to be
\begin{align}
	\op Q_n=\frac{1}{\sqrt{2}}(\op A_n+\op A_n^\dagger), \;\;\;\; \op P_n=\frac{\ii}{\sqrt{2}}(\op A_n^\dagger - \op A_n).
\end{align}
Similarly, for the ladder operators $\{\op a_m, \op a_m^\dagger\}$ and $\{\op{\bar a}_m, \op{\bar a}_m^\dagger\}$ of the local modes we have
\begin{align}
	\op q_m=\frac{1}{\sqrt{2}}(\op a_m+\op a_m^\dagger), \;\;\;\; \op p_m=\frac{\ii}{\sqrt{2}}(\op a_m^\dagger - \op a_m),    \nonumber \\
	\op{\bar q}_m=\frac{1}{\sqrt{2}}(\bar{\op a}_m+\op{\bar a}_m^\dagger), \;\;\;\; \op{\bar p}_m=\frac{\ii}{\sqrt{2}}(\op{\bar a}_m^\dagger - \op{\bar a}_m).
\end{align}
For notational convenience let us define the phase-space vectors $\mat{\op X}=(\op Q_1, \op P_1, \op Q_2, \op P_2, \cdots)^T$, $\mat{\op x}=(\op q_1, \op p_1, \op q_2, \op q_2, \cdots)^T$, and $\mat{\op{\bar x}}=(\op{\bar q}_1,\op{\bar p}_1,\op{\bar q}_2,\op{\bar p}_2,\cdots)^T$.

Within this representation it is straightforward to see that the Bogoliubov transformation from global to local modes, as given in Eq. (\ref{ladderTrans}), is given by the matrix transformations
\begin{align}
	\mat{\op x}=\mat S \mat{\op X}, \;\;\;\;\;\;\;  \mat{\op{\bar x}}=\mat{\bar S} \mat{\op X}, \label{eq:Smatrix1}
\end{align}
where the matrix $\mat S$ takes the block form
\begin{align}  \label{symbogo}
	\mat S= \begin{pmatrix}
		\mat S_{11} & \mat S_{12} & \cdots \\
		\mat S_{21} & \mat S_{22} & \cdots \\
		\vdots & \vdots & \ddots 
	\end{pmatrix},
\end{align}
with 
\begin{align}   \label{symbogo2}
	\mat S_{m n}=\begin{pmatrix}
		\text{Re}(\alpha_{m n}-\beta_{m n}) & \text{Im}(\alpha_{m n}+\beta_{m n}) \\
		-\text{Im}(\alpha_{m n}-\beta_{m n}) & \text{Re}(\alpha_{m n}+\beta_{m n})
	\end{pmatrix}, 
\end{align}
and similarly for $\mat{\bar S}$ using the barred Bogoliubov coefficients. It is straightforward to show that such a transformation preserves the canonical commutation relations iff the Bogoliubov conditions Eq. (\ref{Wron1},\ref{Wron2}) are satisfied.

Using the specific transformation for our scenario, Eq. (\ref{bogos}), we find the $2\times 2$ blocks of matrices $\mat S$ and $\mat{\bar S}$ to be
\begin{equation}
	\! \!\!\! \mat S_{m n}=2 \mathcal V_{mn} \begin{pmatrix}
		\omega_m   \ & 0 \\
		0 & \Omega_n \end{pmatrix}\!\! , \ \  \bar{\mat S}_{m n}=2  \bar{\mathcal{V}}_{mn} \begin{pmatrix}
		\bar \omega_m   \ & 0 \\
		0 & \Omega_n \end{pmatrix} 	 
\end{equation}
We note that the off-diagonal entries of these blocks are zero, resulting from the fact that our Bogoliubov coefficients are purely real. This means that the transformation does not mix canonical position and momentum operators, rather the $\op q$ operators of the local basis are combinations of the global $\op Q$'s only, and similarly for the momentum operators.

It is important to keep in mind that individually the matrices $\mat S$ and $\mat{\bar S}$ are not symplectic. This is because individually they only map onto a subspace of the total Hilbert space of the field. \footnote{The definition of a symplectic matrix $\mat S$ requires that it be square. However if a linear phase space transformation is not square it is still required to preserve the canonical commutation relations. That is, if we have an $m\times n$ transformation matrix $\mat S$ on phase space then it must still satisfy $\mat S \mat \Omega_n \mat S^T=\mat \Omega_m$, where $\mat \Omega_n$ is the $n$-mode symplectic form. If $n>m$ then such a transformation corresponds to a symplectic transformation followed by a partial trace, which can of course bring a pure state to a mixed one.} This is easily concluded from the fact that the reduced field states of the subregions of the cavity are mixed states, despite the global state being pure (the vacuum). A proper symplectic transformation in phase space can always be associated with a unitary operation acting in the Hilbert space, which will always bring a pure state to another pure state.

Rather, it is the combined transformation
\begin{align}    \label{fullSymp1}
	\mat S_\text{Bogo}=\begin{pmatrix}
		\mat S \\
		\mat{\bar S}
	\end{pmatrix}
\end{align}
that is formally symplectic (see the discussion in Sect. \ref{mixednessproblem}). This matrix transforms the global mode basis to the local mode basis, including both sides of the cavity:
\begin{align}
	\begin{pmatrix}
		\mat{\op x} \\
		\mat{\op{\bar x}}
	\end{pmatrix} =\mat S_\text{Bogo} \mat{\op X}.
\end{align}

Given all of this, we are ready to transform the state itself. The global vacuum $\ket{0}_G$ is an example of a Gaussian state, which means that the state is \emph{fully characterized} by its covariance matrix (see Appendix \ref{Gauss}). We will label $\mat \sigma_G$ the covariance matrix of the global vacuum, represented in the global-mode basis. This is simply given by the identity: $\mat \sigma_G=\mat I$. To Bogoliubov transform this state to the local basis, $\mat \sigma_\text{loc}$, we apply the above symplectic transformation to $\mat \sigma_G$: 
\begin{align}   \label{Sigloc}
	\mat \sigma_\text{loc}=\mat S_\text{Bogo} \mat \sigma_G \mat S_\text{Bogo}^T &\equiv
	\begin{pmatrix}
		\mat \sigma && \mat \gamma \\
		\mat \gamma^T && \mat{\bar \sigma}
	\end{pmatrix}   \nonumber \\
	&=\begin{pmatrix}
		\mat S \mat \sigma_G \mat S^T  & \mat S \mat \sigma_G \mat{\bar S}^T  \\
		\mat{\bar S} \mat \sigma_G \mat S^T  & \mat{\bar S} \mat \sigma_G \mat{\bar S}^T
	\end{pmatrix}.
\end{align}
Here the covariance matrix $\mat \sigma = \mat S \mat \sigma_G \mat S^T=\mat S \mat S^T$ represents the reduced field state for the left side of the cavity. Similarly, $\mat{\bar \sigma}=\mat{\bar S} \mat{\bar S}^T$ fully characterizes the reduced state of the right side. The off-diagonal matrix $\mat \gamma=\mat S \mat{\bar S}^T$, on the other hand, contains the correlation structure between the two sides of the cavity.

These matrices are easily computed. We see that each can be split into $2\times 2$ blocks, for example the reduced state of the left side takes the form
\begin{align}  \label{covBlock}
	\mat \sigma= \begin{pmatrix}
		\mat \sigma_{11} & \mat \sigma_{12} & \cdots \\
		\mat \sigma_{21} & \mat \sigma_{22} & \cdots \\
		\vdots & \vdots & \ddots 
	\end{pmatrix}.
\end{align}
Here the $2\times 2$ block $\mat \sigma_{mm}$ is the covariance matrix (i.e. it \emph{is} the reduced state) of the $m$'th local (left side) mode. The off-diagonal block $\mat \sigma_{mn}$, where $m \neq n$, contains the correlations between local modes $m$ and $n$. Using the fact that the $\mat S_{mn}$ are symmetric we see that these blocks are given by $\mat \sigma_{mn}=\sum_\ell \mat S_{m \ell} \mat S_{n \ell}$. Similarly, the state $\mat{\bar \sigma}$ and the correlation matrix $\mat \gamma$ can be split into $2\times 2$ blocks that are given by $\mat{\bar \sigma}_{mn}=\sum_\ell \mat{\bar S}_{m \ell} \mat{\bar S}_{n \ell}$ and $\mat \gamma_{mn}=\sum_\ell \mat S_{m \ell} \mat{\bar S}_{n \ell}$, respectively.

\begin{widetext} These are given by
	\begin{align} \label{covmatR}
	\mat \sigma_{mn}  
	&=\sum_{l} 4 \mathcal{V}_{ml} \mathcal{V}_{nl}
	\begin{pmatrix}
		\omega_m \omega_n & 0 \\     
		0 & \Omega_l^2
	\end{pmatrix}   ,\quad    
		\bar{\mat \sigma}_{mn}  
	&=\sum_{n} 4 \bar{\mathcal V}_{ml} \bar{\mathcal V}_{nl}
	\begin{pmatrix}
		\bar {\omega}_m \bar \omega_n & 0 \\     
		0 & \Omega_l^2
	\end{pmatrix}    ,\quad   
	\mat \gamma_{mn}
	&=\sum_{l} 4 \mathcal V_{ml} \bar{\mathcal V}_{nl}
	\begin{pmatrix}
		\omega_m \bar{\omega}_n & 0 \\     
		0 & \Omega_l^2
	\end{pmatrix}.
	\end{align}
\end{widetext}

Together, these blocks constitute a full characterization of the global vacuum in the local-mode basis, and in particular $\mat \sigma$ fully characterizes the reduced state of the left side of the cavity. Although we have derived the full analytical expressions, it should be noted that in the remainder of the paper, when we present quantitative results, we have done so by computing the above matrix elements numerically, by performing the sums to convergence.

There are several observations that we can make from this result. The first is that the reduced states $\mat \sigma$ and $\mat{\bar \sigma}$ are clearly excited states, meaning in this language that they are not equal to the identity (the vacuum). Mathematically, this is what is meant by the statement ``half of an empty box is non-empty". Equivalently, this is a mathematical description of the particle creation due to instantaneously slamming down a mirror. Another observation is that the correlation structure of the global vacuum in this basis is extremely connected, meaning that every local mode is correlated (if perhaps not entangled) with every other local mode. That is, since the blocks $\mat \gamma_{mn}$ are nonzero this means that every local mode of the left side is correlated with every local mode of the right, and vice versa. Similarly, every local mode is correlated with every other local mode of the same side, as demonstrated by the fact that the blocks $\mat \sigma_{mn}$ and $\mat{\bar \sigma}_{mn}$ are nonzero.

\subsection{The state of three regions}    \label{stateThree}

We will now outline exactly the same procedure for the case of three regions in the cavity (equivalently, the case where two mirrors are simultaneously introduced).  This will allow us to consider the entanglement between spatially-separated regions (i.e. the leftmost and rightmost regions). As we will see, this is crucial for demonstrating that the entanglement obtained by slamming mirrors is derived from the previously existing vacuum entanglement, rather than having been created by the slamming process. 

The procedure follows from the Bogoliubov transformation described in Sect. \ref{three1}. We will also describe how to obtain the reduced state of two out of the three regions (in fact this is trivial in the language of covariance matrices). In the phase space representation we have equivalent matrix equations as those in Eq. (\ref{eq:Smatrix1},\ref{symbogo},\ref{symbogo2}), i. e.:
\begin{align}
	\mat{\op x}^Z=\mat S^Z \mat{\op X},  \qquad Z=A,B,C
\end{align}
where $\mat S^Z$ has the block form as given in Eq. (\ref{symbogo}): 
\begin{align}   
	\mat S^Z_{m n} &=\begin{pmatrix}
		\text{Re}(\alpha^Z_{m n}-\beta^Z_{m n}) & \text{Im}(\alpha^Z_{m n}+\beta^Z_{m n}) \\
		-\text{Im}(\alpha^Z_{m n}-\beta^Z_{m n}) & \text{Re}(\alpha^Z_{m n}+\beta^Z_{m n})
	\end{pmatrix}\\
	&=2 \mathcal V_{mn}^Z \begin{pmatrix}
		\omega^Z_m   \ & 0 \\
		0 & \Omega_n
	\end{pmatrix}.
\end{align}

The combined transformation, that which is formally symplectic, is given in analogy to Eq. (\ref{fullSymp1}):
\begin{align}
	\mat S_\text{Bogo}=\begin{pmatrix}
		\mat S^A \\
		\mat S^B \\
		\mat S^C 
	\end{pmatrix},
\end{align}
and transforms the global mode basis to the local mode basis of the three regions:
\begin{align}
	\begin{pmatrix}
		\mat{\op{x}}^A \\
		\mat{\op{x}}^B \\
		\mat{\op{x}}^C
	\end{pmatrix} =\mat S_\text{Bogo} \mat{\op X}.
\end{align}

Again, to Bogoliubov transform the global state $\mat \sigma_G = \mat I$ to the local basis we apply this transformation to $\mat \sigma_G$: 
\begin{align}   \label{threeSig}
	\mat \sigma_\text{loc} &=\mat S_\text{Bogo} \mat \sigma_G \mat S_\text{Bogo}^T \equiv
	\begin{pmatrix}
		\mat \sigma_A && \mat \gamma_{AB}   && \mat \gamma_{AC}\\
		\mat \gamma^T_{AB} && \mat{ \sigma}_B && \mat \gamma_{BC}\\
		\mat \gamma^T_{AC} && \mat \gamma^T_{BC}   && \mat \sigma_{C}
	\end{pmatrix}   \nonumber \\
	&=	\begin{pmatrix}
		\mat S^A \mat \sigma_G \mat S^{A^T} && \mat S^A \mat \sigma_G \mat S^{B^T}    && \mat S^A \mat \sigma_G \mat S^{C^T} \\
		\mat S^B \mat \sigma_G \mat S^{A^T}  &&\mat S^B \mat \sigma_G \mat S^{B^T} && \mat S^B \mat \sigma_G \mat S^{C^T} \\
		\mat S^C \mat \sigma_G \mat S^{A^T}  && \mat S^C \mat \sigma_G \mat S^{B^T} && \mat S^C \mat \sigma_G \mat S^{C^T} 
	\end{pmatrix}.
\end{align}
The blocks again represent the reduced state of, and the correlations between, the three regions. For example $\mat \sigma_A$ is the reduced state of the left-most region and $\mat \gamma_{AC}$ contains the correlations between the left-most and right-most regions. As before, each matrix can be further split into 2x2 blocks given by $\mat \sigma^Z_{mn}=\sum_\ell \mat S^Z_{m \ell} \mat S^Z_{n \ell}$, $\gamma^{YZ}_{mn}=\sum_\ell \mat S^Y_{m \ell} \mat S^Z_{n \ell}$. These are given by

	\begin{align} 
	\mat \sigma^{Z}_{mn}  
	&=\sum_{l} 4 \mathcal V_{ml}^Z \mathcal V_{nl}^Z
	\begin{pmatrix}
		\omega^{Z}_m \omega^{Z}_n & 0 \\     
		0 & \Omega_l^2
	\end{pmatrix}      \\
	\mat \gamma^{YZ}_{mn}
	&=\sum_{l} 4 \mathcal V_{ml}^Y \mathcal V_{nl}^Z
	\begin{pmatrix}
		\omega^{Y}_m \omega^{Z}_n & 0 \\     
		0 & \Omega_l^2
	\end{pmatrix}.
	\end{align}

From here, one may easily study the reduced state of two of the three regions by simply taking the appropriate blocks of Eq. \ref{threeSig}. For example the reduced state of system $AC$ (the left-most and right-most regions) is obtained by tracing out $B$, which here simply results in the covariance matrix
\begin{align}    \label{ACstate}
	\mat \sigma_\text{AC} &=
	\begin{pmatrix}
		\mat \sigma_A &&   \mat \gamma_{AC}\\
		\mat \gamma^T_{AC}   && \mat \sigma_{C}
	\end{pmatrix}   \nonumber \\
	&=	\begin{pmatrix}
		\mat S^A \mat \sigma_G \mat S^{A^T}   && \mat S^A \mat \sigma_G \mat S^{C^T} \\
		\mat S^C \mat \sigma_G \mat S^{A^T}  && \mat S^C \mat \sigma_G \mat S^{C^T} 
	\end{pmatrix}.
\end{align}

\section{With vs. without a mirror}

Before we proceed to analyse other local features like the entanglement between left and right regions of the cavity, we would like to make a stop to discuss a little bit more conceptually the differences between the analysis of the two possible scenarios, with and without introducing the mirror. Again, what does it mean for half of an empty box to be non-empty?  We know that in some sense the reduced state of a subregion of the global vacuum is excited; certainly the state $\mat \sigma$ in Eq. (\ref{covmatR}) is an excited state (that is, excited with respect to the local-mode basis, which is the whole point). However, what does this mathematical fact have to do with reality? As discussed earlier, the answer, in fact, is that the real excitations produced by the mirror are mathematically equivalent to the virtual local excitations attributed to vacuum entanglement. Their measurement, therefore, constitutes an achievement of our goal.

\subsection{Time evolution}   \label{timeevolution}

Both of the scenarios, with and without a mirror, are equivalent at time $t=0$. This implies that the Bogoliubov transformation will be exactly the same for both sets (primed and unprimed modes as discussed in the previous sections) as the transformation coefficients are computed using the Klein-Gordon inner product, which contains only the mode functions and their first time-derivatives \cite{BirrelDavies84}). Thus, the field state of the left-cavity immediately following the introduction of the mirror will, in fact, be given exactly by the covariance matrix $\mat \sigma$ as given by Eqs. (\ref{covBlock},\ref{covmatR}). The only difference now is that the mode-basis that $\mat \sigma$ is associated with is different, in the sense that it evolves differently for $t>0$. Similarly the reduced state of the right-cavity will be given by $\mat{\bar \sigma}$ and the correlations between the two (separated) cavities will be contained in $\mat \gamma$, the blocks of each being given by Eqs. (\ref{covmatR}). Importantly, this means that the entanglement structure contained in the state is exactly the same  in both cases. That is, the real particles created in the left-side by slamming down a mirror are entangled with the created particles in the right-side, and this entanglement has exactly the same structure that the original vacuum entanglement present before the mirror was introduced. We will fully discuss this entanglement in Sect. \ref{sectEnt}.

But surely the state of the field has been changed due to the introduction of the mirror. Clearly in some sense it has. We have created real particles. We have added energy to the system by changing the Hamiltonian. The state of the new left-side cavity (for example) is certainly time-dependent. This is not surprising, as we would expect a burst of particles to be propagating away from the newly introduced mirror (shortly we will discuss this further). The reduced state of the left-side of the larger cavity (without a mirror), on the other hand, is by construction time-independent. The global vacuum $\ket{0}_G$ is a stationary state with respect to the global Hamiltonian $\op H_G$, and thus the reduced state will be time-independent as well. In this sense the two states are certainly different.

Nevertheless the state at $t=0$ is described by exactly the same covariance matrix. We will now elucidate the nature of time evolution in the case that a mirror has been slammed; indeed we will take advantage of a subtlety in the time evolution that is particularly apparent when working with covariance matrices. First consider, for example, working in the Schr\"odinger picture. In this case the field in the left-cavity is time-independent: $\op \phi(x,t)=\op \phi(x,t=0)=\sum_m (u_m(x,0) \op a_m+u^*_m(x,0) \op a_m^\dagger)$, where a $u_m(x,0)=\frac{1}{r\omega_m}\sin \frac{\pi m x}{r}$. The state $\op \rho(t)$ is what evolves, and this gives a corresponding time evolution to the covariance matrix elements via $\sigma_{mn}(t)=\text{Tr}(\op \rho(t)(\op x_m \op x_n + \op x_n \op x_m))$. This free evolution can be represented symplectically: $\mat \sigma(t)=\mat S_F(t) \mat \sigma \mat S_F(t)^T$ where \cite{Brown13}
\begin{align}   \label{freeEv}
	\mat S_F(t)=\bigoplus_m \begin{pmatrix}
		\cos \omega_m t & \sin \omega_m t \\
		- \sin \omega_m t & \cos \omega_m t
	\end{pmatrix}.
\end{align}
Alternatively we can work in the Heisenberg picture, in which the field is the time-dependent operator
\begin{align}  \label{field3}
	\op \phi(x,t)&=\sum_m (u_m(x,0) \op a_m e^{-\ii \omega_m t}+u^*_m(x,0)\op a_m^\dagger e^{\ii \omega_m t})   \nonumber \\
	&=\sum_m (u_m(x,t) \op a_m+u_m^*(x,t)\op a_m^\dagger).
\end{align}

A subtle issue, however, is that the Heisenberg evolution of the field can be viewed in two ways, as given by the two lines above. In the first line it is the operators themselves that evolve: $\op a_m(t)=\op a_m e^{-\ii \omega_m t}$. This corresponds to an evolution of the quadrature operators $\op x_m(t)$ that leads to a symplectic evolution $\mat S_F(t)$ of the covariance matrix, equivalent to what was obtained in the Schr\"odinger picture. A key observation is that in both of these pictures it is the \emph{time-independent mode-functions} $u_m(x,0)$ that the \emph{time-dependent covariance matrix} $\mat \sigma(t)$ is associated with. The other way of viewing the evolution, as indicated by the second line in Eq. (\ref{field3}), is to keep the operators themselves time-independent (thus giving a time-independent $\mat \sigma$) and to rather let the mode-functions $u(x,t)$ contain the time evolution. In this case the covariance matrix does not change, but it is understood that the mode-functions with which it is associated \emph{do} evolve. 

This last picture is the one that we will adopt here, in all work below. In this way we do not need to actually consider any evolution in the covariance matrix directly; our state will always be described by the matrix $\mat \sigma$, the same one used to describe the spatial reduced state in the case without a mirror. The time-evolution induced by slamming a mirror is then trivial: it is simply given by the time-dependence already present in the $t>0$ mode functions defined within the left cavity as $u_m(x,t)=u_m(x,0)e^{-\ii \omega_m t}$ and within the right cavity as $\bar u_m(x,t)=\bar u_m(x,0)e^{-\ii \bar \omega_m t}$

\subsection{Finite-time mirror}  \label{finitetime}

In the calculations of the next section we will continue to assume an instantaneous introduction of the mirror(s) in the cavity. Before devoting ourselves to this, however, we should briefly discuss how the physics changes if the introduction of the mirror takes place within a finite time window $\Delta t$, as of course will always be the case in any physical realization. Let us continue to assume that the introduction happens very fast as compared to the fundamental time scales of the reduced cavities: $\Delta t \ll 1/\omega_1$ and $\Delta t \ll 1/\bar{\omega}_1$. In this case the low-energy local modes will still see the mirror appear very quickly (i.e. as compared to their free evolution time scale), and so their reduced states and correlations amongst themselves will be well approximated by the covariance matrices of Eqs. (\ref{covmatR}). That is, within a low energy sector (the limit of which is determined by how fast the mirror can be introduced) the results that we will present will hold to a good approximation. On the other hand for the very high-energy modes (that see the introduction of the mirror occur very slowly) we can make an adiabatic approximation to conclude that they will evolve to their local ground states. That is, if $m$ is large enough such that $\Delta t \gg 1/\omega_m$ then after the cavity is introduced the reduced state of this mode will approximately be $\ket{0}_m$, defined to satisfy $\op a_m \ket{0}_m=0$, and will have vanishing correlation with the rest of the system. Clearly there will be a smooth transition between these two regimes, which our work does not capture. Nevertheless by considering only a finite number of modes $N$, as we will be doing, our description of this set will be accurate as long as $\Delta t \ll 1/\omega_N$. 

Note also that, in terms of application, the amount of entanglement that one obtains between cavities after slamming a mirror (which we will discuss in the next section) depends on how fast one's mirror is slammed. The faster it can be achieved, the more entanglement will remain in the two cavities afterwards. This is because the high-energy modes contain entanglement, and thus the more of these modes whose states are not significantly altered by introducing the mirror, the more entanglement we will retained. For modes of too-high energy, $\Delta t \gg 1/\omega_m$, the act of slamming the mirror will destroy the correlations that they have with the opposite side of the cavity.


\section{Entanglement}    \label{sectEnt}

We will now enter the results section of our paper. We will discuss various aspects of entanglement between the two sides of the cavity (equivalent in both the cases of with and without a cavity, as discussed above). As part of our exposition we will propose a spatial distribution of entanglement between the two sides of the cavity, and see how this naturally leads to the physical picture of bursts of (entangled) particles being produced by slamming down a mirror. We will begin by just discussing a single mirror, and later will move on to the two-mirror case. We will show that with two mirrors, slammed simultaneously some distance apart, there is still entanglement retained between separated regions (i.e. left-most and right-most). We will also discuss how the act of slamming down a mirror can be interpreted as an efficient method of vacuum entanglement harvesting.

Our result are computed numerically from the covariance matrices presented Sect. \ref{Computestate}. To do so, however, we must restrict ourselves to finite matrices. This means taking only a finite number of local modes $N$, both on the left and right sides. That is, what we actually consider is the reduced state of the first $N$ local modes on each side. This is actually not physically unrealistic since, as discussed in Sect. \ref{finitetime}, our analysis will only be valid for some low-energy regime anyway, depending on how fast the mirror is slammed. Numerically, unless otherwise stated we will take $N=200$. Note, however, that the reduced state of these first $N$ local modes is exact up to numerically negligible addends. That is, in performing the Bogoliubov transform we made sure to include enough global modes in the sum of Eq. (\ref{covmatR}), such that our results converge.

\subsection{Mode-mode entanglement}

With the state of the global vacuum represented in the local-mode basis, as given by Eqs. (\ref{covmatR}), we can characterize the entanglement between the two sides of the cavity. We can, for example, consider the two-mode entanglement between each pair of local modes on the left and right side. The correlations between each pair (as given by the two-point correlators of the number operators) have already been computed in \cite{Vazquez14}. However,  for each two-mode pair the fact that they are correlated does not imply that they are entangled because the two-mode state of this pair is mixed. Thus, to extend upon the results of \cite{Vazquez14} we compute the logarithmic negativity $E_N$ \cite{Plenio05} of each pair between the two sides. 

\begin{figure*}[t]
	\centering
                 \includegraphics[width=\textwidth]{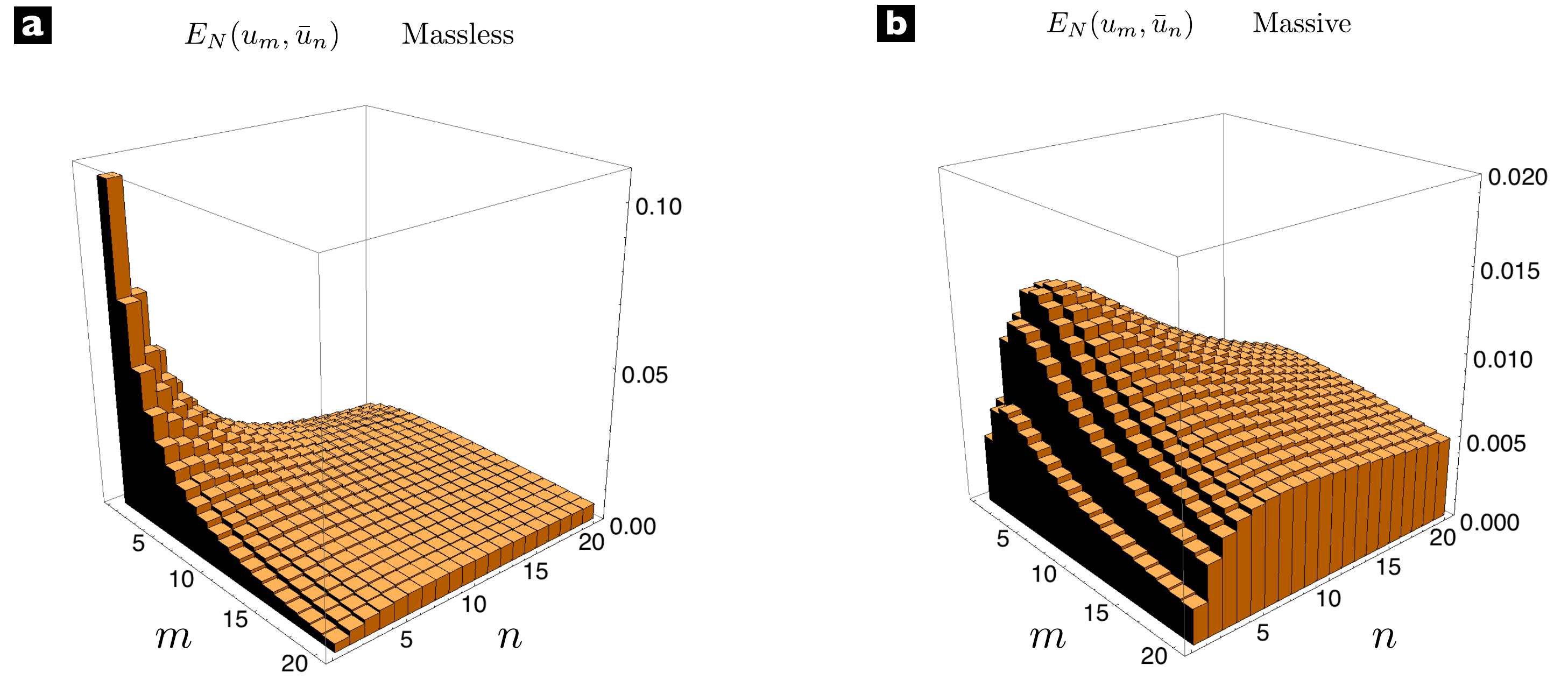}
	\caption{(Color online) The logarithmic negativity $E_N$ between local modes $u_m$ and $\bar u_n$ on the left and right sides of the cavity, respectively. The cavity is divided in two equal regions $r=0.5 R$ . Left: a field mass of $\mu=0$. Right: a field mass of $\mu=15/R$.}
        \label{Eu}
\end{figure*}

To this end, we take the $4\times 4$, two-mode covariance matrix (i.e. the reduced state) of mode $m$ on the left and mode $n$ on the right of the cavity. This is simply
\begin{align}
	\mat \sigma_\text{two mode}=\begin{pmatrix}
		\mat \sigma_{mm} & \mat \gamma_{mn} \\
		\mat \gamma_{mn}^T & \mat{\bar \sigma}_{nn}
	\end{pmatrix}.
\end{align}
From here, we can apply Eq. (\ref{logneg}) to compute $E_N$ between the two modes. The result is displayed in Fig. \ref{Eu}, where we consider field masses $\mu=0$ and $\mu=15/R$ . The cavity is split in two equal regions as $r=0.5 R$. We observe that, perhaps remarkably, nearly every mode is entangled with almost every other. Eventually as $m$ and $n$ become sufficiently different the two-mode entanglement tends to vanish (although they will always have nonzero correlation), but we can see that the decay is very slow. It should be noted that we can similarly compute the entanglement between different local modes from the same side, and in fact doing so produces a qualitatively equivalent plot. A particularly striking feature of the mode-mode entanglement is that the peak entanglement moves to higher mode numbers as the mass of the field is increased. \footnote{This behavior is actually expected from the fact that the correlation length in a field goes as the Compton wavelength \cite{Zych10}, meaning that correlations become more spatially confined with higher mass $\mu$. It follows that what correlations are present between the two sides should be more supported within the modes of smaller wavelength, i.e. those of higher frequency.}
This figure clearly demonstrates that the two sides of the cavity are entangled. Even a single pair with nonzero entanglement demonstrates this. However even if every pair were separable this would \emph{not} constitute a proof that the two sides as a whole are separable. 

This leads to the question of the full, many-mode entanglement between the two sides. We can certainly compute this, given some set of $N$ local modes on either side \cite{Adesso07} (specifically we would compute the negativity, not the reduced entropy, as we explain in a moment). This of course gives a non-zero answer, however it is questionable how useful the numerical answer actually is because it will always depend on the number of local modes $N$ considered. The entanglement increases with $N$, and we expect that it diverges in the $N \rightarrow \infty$ limit (check footnote \ref{footinfty}), given that the vacuum entanglement is typically known to be UV-divergent. We will thus not concern ourselves with this calculation explicitly. Nevertheless there is a related issue that should be discussed before moving on, which we will now focus on.

\subsection{The mixedness problem}   \label{mixednessproblem}

One would assume that in order to compute the entanglement one should simply compute the reduced entropy (as given by Eq. (\ref{entropy})) of one side of the cavity, since the global state is pure. Formally this is true of course, but interestingly the reduced entropy will never be an entanglement measure if one only considers a finite number of local modes $N$, and in fact this can never be remedied by simply increasing $N$.

This occurs because, as we have just seen, there is quite a lot of correlation between local modes of different number. This means that the left-side state $\mat \sigma$ (with finite number of modes) will not just be entangled with the opposite side of the cavity but also with the higher-number modes on the same side. That is, the entropy $S(\mat \sigma)$ is \emph{not} a measure of entanglement with the other side, but rather with the other side plus all of the higher modes that we have traced away. Put another way: if we compute the full state of both sides $\mat \sigma_\text{loc}$, but with the understanding that this corresponds to the reduced state of the first $N$ local modes on the left with the first $N$ on the right (and their correlations), then this state will be mixed despite the fact that the global vacuum is pure. Equivalently, the transformation of Eq.(\ref{fullSymp1}) will never in practice be sympletic. What's more (and rather interestingly) this problem does \emph{not} get better as $N$ is increased \footnote{\label{footinfty}In fact, as we increase the number $N$ of local modes considered (on both sides of the cavity) the global state we obtain becomes \emph{more} mixed, with a higher entropy. We suspect that the entropy diverges in the $N \rightarrow \infty$ limit, despite the fact that in a formal sense the result should be a pure state. After a moment of thought this is actually not overly surprising. Consider for a moment the very different system of a spatial volume in free Minkowski space with a field in the Minkowski vacuum. It is well known that the entropy of the reduced state inside the volume scales as its area, meaning that as this region is expanded it becomes more mixed. Thus, despite the field over all of space being in a pure state, one can never approach this by taking the limit of larger and larger regions (the entropy will diverge as the region expands to infinity). In this example the area-law can be physically understood by taking a spatial discretization of the field. A given spatial degree of freedom will largely only be entangled with its nearest neighbors, and thus the area law can be understood considering that the area is proportional to the number of nearest-neighbor connections that the entangling surface crosses. In our scenario we have seen that the global vacuum has a very densely connected entanglement structure in the local-mode basis. Every local mode is entangled with many others, including many others of higher frequency. Thus, by increasing the number of local modes $N$ that we consider we are increasing the number of ``entanglement connections" between low and high modes that are separated by the cutoff. Given this intuition it makes sense that the entropy of our global state should increase with increasing $N$; it arises as a consequence of the system being highly connected. Even so, it is interesting (and perhaps disconcerting) that in the local-mode basis one can never approach purity by considering more and more modes. We suspect that this is deeply connected to the note made in \cite{Vazquez14} regarding the unitary inequivalence between the global and local mode bases.}.

Importantly, such an issue will never arise in any real scenario of a slamming mirror; a finite slamming time $\Delta t$ fixes this mixedness problem. The introduction of a mirror is just represented by a time-dependent Hamiltonian, and so of course the evolution of the field under this action must be unitary. The system of the two new cavities combined, therefore, must be in a pure state. As discussed in Sect. \ref{finitetime}, a finite $\Delta t$ will mean that local modes of high enough frequency will not actually be in the state nor share the correlations as predicted from the covariance matrices in Sect. \ref{Computestate}, which were computed assuming instantaneous slamming. For a real situation, high-frequency local modes will be nearly in their ground states, and importantly have vanishing correlations with anything else, thus remedying the origin of the mixedness problem. The global state in the local basis will indeed be pure beyond a given energy scale, as it must be.

\subsection{Symplectic diagonalization}    \label{SympDiag}

Here we will describe the process of symplectically diagonalizing the local states $\mat \sigma$ and $\bar{\mat \sigma}$. This is method by which we can greatly simplify the entanglement structure between the two sides which, given the complexity seen in Fig. \ref{Eu}, will be a considerable advantage. We will see in later sections how this process also allows us to make conclusions about the spatial distribution of entanglement as well as see very clearly the propagating ``burst" of particles that is produced by slamming down a mirror.

The specifics of local, symplect diagonalization and the method for finding the correct transformations matrices are described  in Appendices \ref{Gauss} and \ref{diagonalization}. We (numerically) find symplectic matrices $\mat S_D$ and $\mat{\bar S}_D$ that diagonalize $\mat \sigma$ and $\mat{\bar \sigma}$, respectively: $\mat S_D \mat \sigma \mat S_D^T=\mat D$ and $\mat{\bar S}_D \mat{\bar \sigma}\mat{\bar S}_D^T=\mat{\bar D}$ where
\begin{align}
	\mat D=\bigoplus_m \begin{pmatrix}
		\nu_m & 0 \\
		0 & \nu_m
	\end{pmatrix},   \;\;\;\;
	\mat{\bar D}=\bigoplus_m \begin{pmatrix}
		\bar \nu_m & 0 \\
		0 & \bar \nu_m
	\end{pmatrix}.
\end{align}
Here $\nu_m$ and $\bar \nu_m$ are the symplectic eigenvalues of $\mat \sigma$ and $\mat{\bar \sigma}$, respectively. Let's just consider the left side for a moment: $\mat \sigma \rightarrow \mat D$. This is simply a change of mode-basis, and we can compute the mode functions associated with this new basis by reading off the Bogoliubov coefficients from $\mat S_D$ via reversing Eqs. (\ref{symbogo},\ref{symbogo2}) . Here we will label these coefficients $\zeta_{\ell m}$ and $\eta_{\ell m}$ (in place of the usual $\alpha$ and $\beta$ notation, respectively). These new mode functions, which we will label $v_\ell(x,t)$, are thus given by
\begin{align}  \label{UtoV}
	v_\ell(x,t)&=\sum_m (\zeta_{\ell m}u_m(x,t)+\eta_{\ell m}u_m^*(x,t))  \nonumber \\
	&=\sum_m \frac{1}{\sqrt{r \omega_m}}\sin \left(\frac{\pi m x}{r}\right)(\zeta_{\ell m}e^{-\ii \omega_m t}+\eta_{\ell m}e^{\ii \omega_m t}).
\end{align}
We can similarly define a new set of local modes $\bar v_\ell(x)$ on the right side of the cavity.

We remind the reader that (as discussed in Sect. \ref{timeevolution}) we are working in the ``Heisenberg picture", but not such that the $\op q$ and $\op p$ operators evolve (i.e. our covariance matrix is time-independent) but rather such that the mode functions with respect to which we represent the state themselves evolve. In this picture the diagonalizing transformations are of course time-independent (since the covariance matrix is time-independent). We could, however, arrive at the same set of $v$-modes working directly in the Schr\"odinger picture, in which the diagonalizing transformation would be time-dependent \footnote{This can also be done in either of the pictures in which it is the covariance matrix that evolves, $\mat \sigma(t)=\mat S_F(t) \mat \sigma \mat S_F(t)^T$, and in which the spatial modes are time independent, $u_m(x,0)$. In this case the diagonalizing transformation will be time-dependent: $\mat S_D(t)$. However the symplectic spectrum of $\mat \sigma(t)$ will be time-independent, being symplectically invariant. Thus we have $\mat D=\mat S_D \mat \sigma \mat S_D^T=\mat S_D(t) \mat \sigma(t)\mat S_D(t)^T=\mat S_D(t) \mat S_F(t) \mat \sigma \mat S_F(t)^T \mat S_D(t)^T$, from which we can represent the time-dependent diagonalizing transformation as $\mat S_D(t)=\mat S_D \mat S_F(-t)$. We can use this to compute the corresponding time-dependent Bogoliubov coefficients $\gamma_{\ell m}(t)$ and $\eta_{\ell m}(t)$. Using Eq. (\ref{freeEv}) and the relation between a symplectic transformation  and its corresponding Bogoliubov coefficients, as given by Eqs. (\ref{symbogo},\ref{symbogo2}), it is straightforward to find that $\gamma_{\ell m}(t)=\gamma_{\ell m}e^{-\ii \omega_m t}$ and $\eta_{\ell m}(t)=\eta_{\ell m}e^{\ii \omega_m t}$, in agreement with Eq. (\ref{UtoV}).}.

This is a change of mode basis which results in all left-side modes $v_\ell(x,t)$ being in a product state with respect to each other, and similarly with the right-side modes $\bar v(x,t)$. I.e. the transformation $\mat S_D$ removes all correlations between modes on the left side. In this way we are isolating exactly the local spatial modes that contain the entanglement between $\mat \sigma$ and the rest of the system. Furthermore, it turns out that in our system the first mode in this new basis, the one associated with symplectic eigenvalue $\nu_1$ and spatial mode $v_1(x,t)$, is the mode that contains the large majority of the mixedness in $\mat \sigma$. That is, almost all of the symplectic eigenvalues have values very near to unity, meaning that the corresponding modes are very nearly pure. The first value, $\nu_1$, is by far the largest. For example with the parameters $r=0.5 R$, $\mu=0$, and $N=200$ (the number of local modes considered) the first several symplectic eigenvalues take the values $\{\nu_\ell\}=(1.840, 1.051, 1.004, 1.000, \cdots)$. Note that as $N$ is increased these values (and thus the entropy of $\mat \sigma$) increase as well. All of this applies equally well to the right-side transformation $\mat{\bar \sigma}\rightarrow \mat{\bar D}$ via $\mat{\bar S}_D$.

\begin{figure*}[t]
	\centering
                 \includegraphics[width=\textwidth]{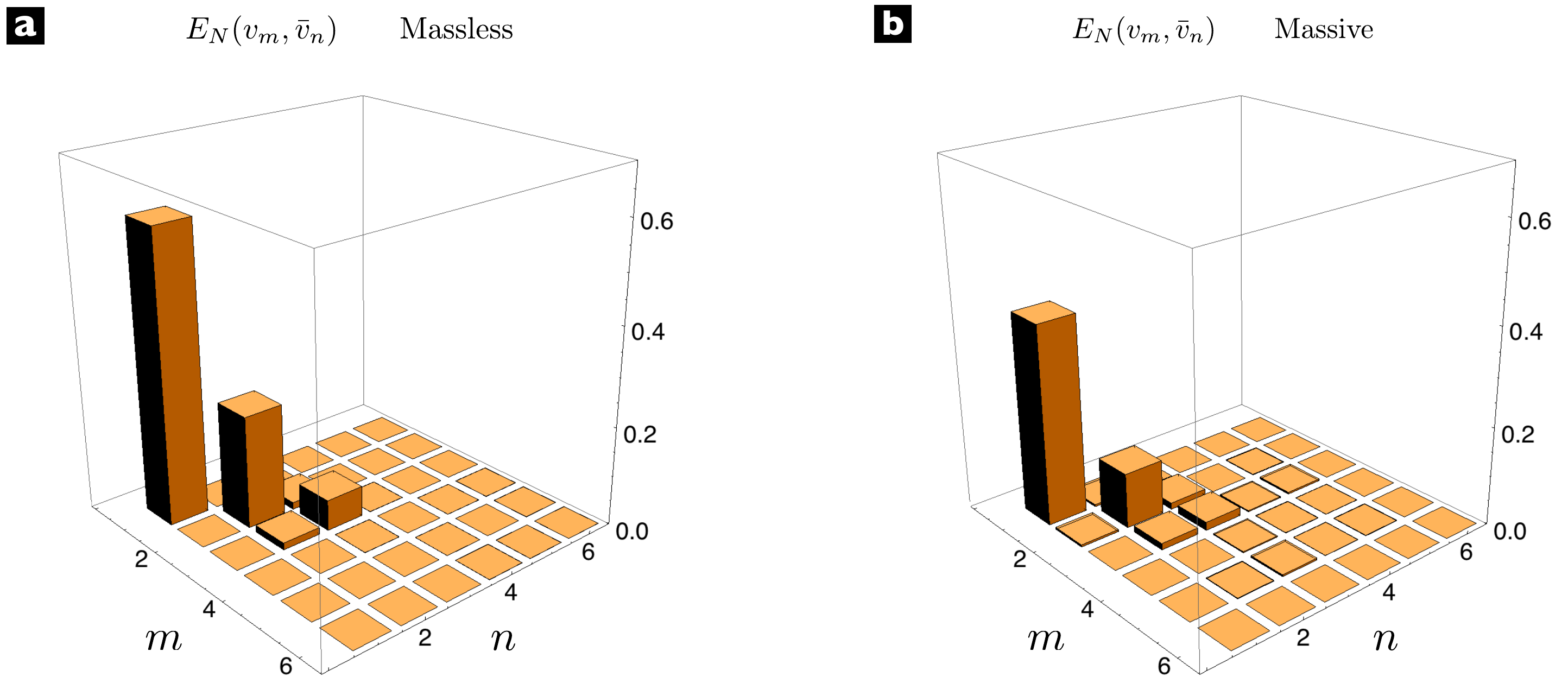}
	\caption{(Color online) The logarithmic negativity $E_N$ between local, diagonalizing modes $v_m$ and $\bar v_n$ on the left and right sides of the cavity, respectively. The cavity is split into two equal sides, $r=0.5R$, and $N=200$ for both the left and right sides.  Left: a field mass of $\mu=0$. Right: a field mass of $\mu=15/R$.}
        \label{Ev}
\end{figure*}

As elaborated on in Appendix \ref{Gauss}, if the state $\mat \sigma_\text{loc}$ of both sides were pure then applying the local transformation $\mat S_D \oplus \mat{\bar S}_D$ to $\mat \sigma_\text{loc}$ would also diagonalize the off-diagonal (correlation) block $\mat \gamma$. Were this the case then the local mode $v_1(x,t)$ on the left side would be solely correlated with the corresponding mode $\bar v_1(x,t)$ on the right side, and similarly for the higher $v$-modes. Unfortunately, as discussed above, when taking a finite $N$ we necessarily find that $\mat \sigma_\text{loc}$ is a mixed state. This means that a local symplectic diagonalization does not produce this one-to-one correspondence between the two sides. Despite this, however, we have found that in fact we very nearly do obtain this correspondence upon local diagonalization. This can be seen in Fig. \ref{Ev} where we plot the logarithmic negativity between modes $v_\ell(x,t)$ and $\bar{v}_\ell(x,t)$ similarly to what is plotted in Fig. \ref{Eu} for the $u$-modes. Here we have taken $N=200$ for both the left and right sides. We see that indeed, despite $\mat{\sigma}_\text{loc}$ being mixed, the majority of the entanglement between the two sides is contained in $v_1(x,t)$ and $\bar v_1(x,t)$ (we could also plot the mutual information between modes, in order to get a better idea of the correlations in general, but the result looks nearly identical qualitatively).

\begin{figure}[t]
	\centering
    \includegraphics[width=0.48\textwidth]{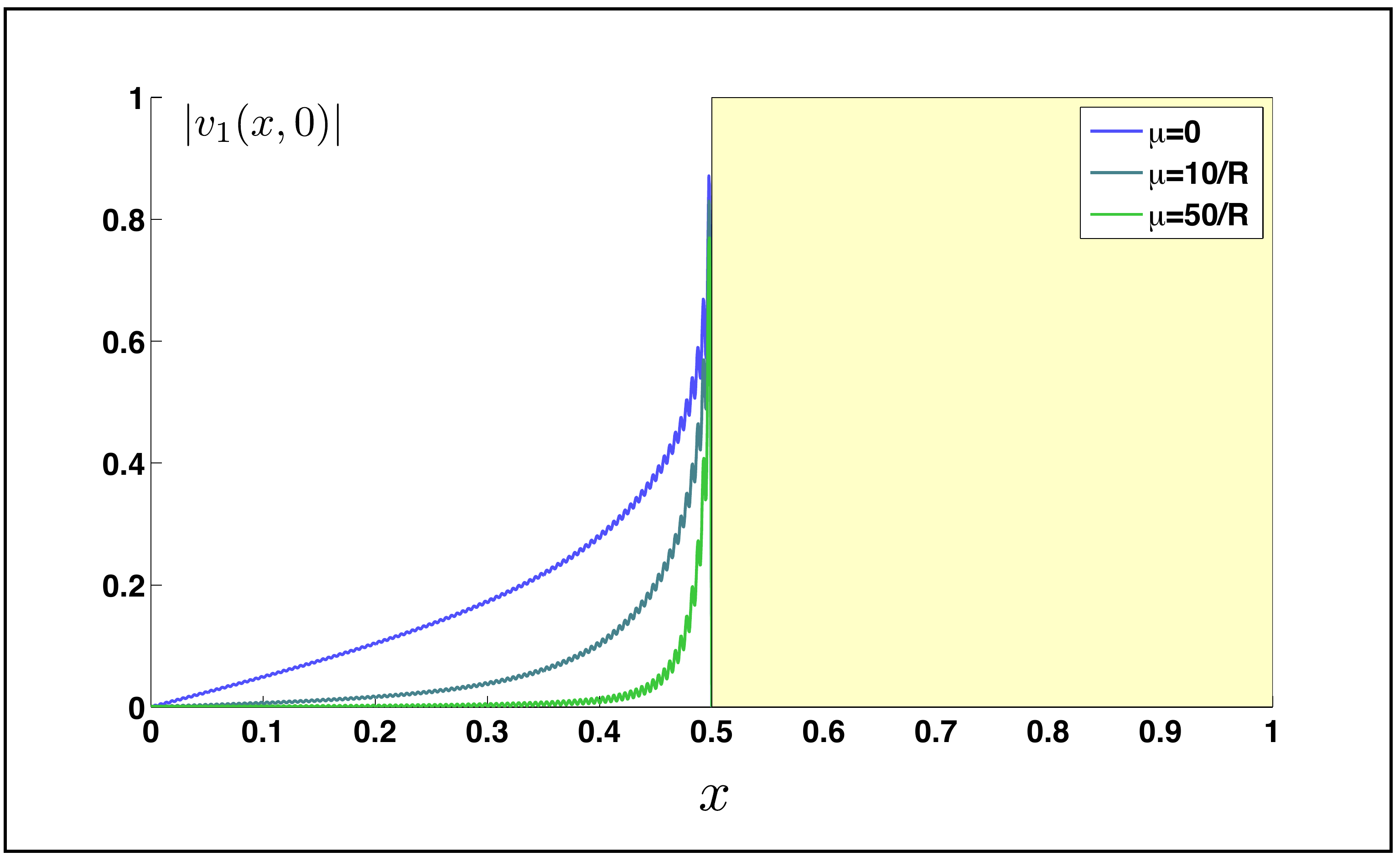}
	\caption{(Color online) The function $|v_1(x)|$ in the left-side of the cavity, representing the spatial distribution of entanglement with the opposite side. The parameters are given by $r=0.5R$, and $N=200$, with different field masses $\mu$ considered: $0$ (blue), $10/R$ (light blue) and $50/R$ (green). As can be seen: the larger the mass of the field, the closer the entanglement straddles the boundary between the two sides of the cavity, as expected. }
        \label{v1}
\end{figure}

\begin{figure}[t]
	\centering
    \includegraphics[width=0.48\textwidth]{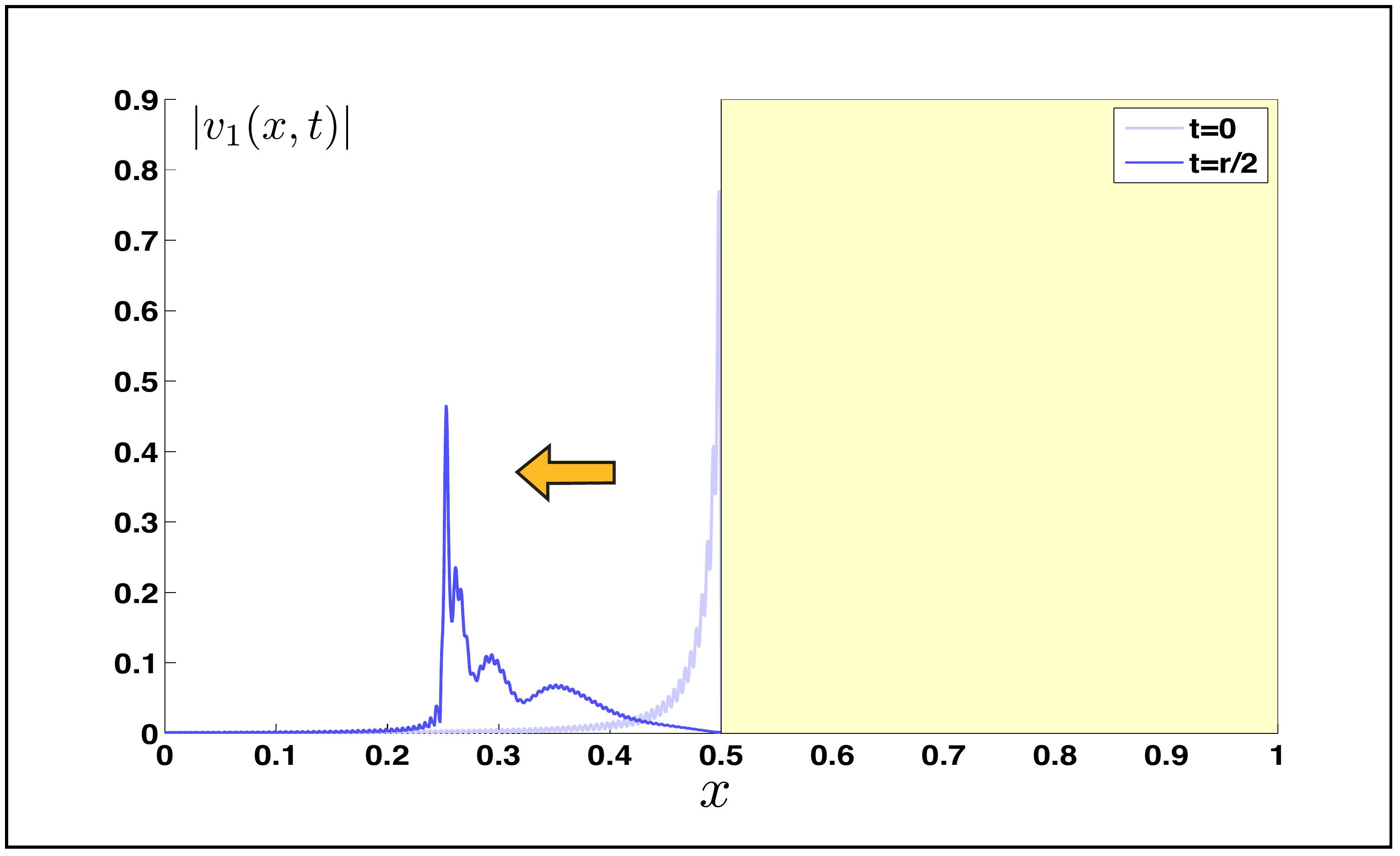}
	\caption{(Color online) Evolution of the entanglement spatial distribution for the massive case $\mu=50/R$ after an elapsed time $t = r/2$. We can see a peak for the correlations at exactly the position of the particle-burst front as originated from the slamming. The cavity parameters are the same as in Fig. \ref{v1}}
        \label{burst}
\end{figure}

\subsection{Spatial structure of entanglement}

One immediate application of finding the locally, symplectically diagonalizing basis is that we are able to discuss and make observations about the spatial structure of entanglement between the two sides of the cavity. For this section we will take $t=0$, by which we are discussing the local physics of the cavity before the mirror has been introduced. That is, in this section we are simply asking about the local properties of a vacuum field, and not considering yet the time evolution caused by introducing a mirror.

We know that there is spatial vacuum entanglement; the two sides of the cavity are entangled. This fact alone, however, gives no information on how entanglement is spatially distributed. From what is known about vacuum entanglement we expect it to be spatially focused near the boundary between the two regions, since the correlation in a field decays with distance \cite{Casini09,Cramer06,Amico08}. It is this that leads, for example, to the well-known area-law for the entanglement entropy. There is also evidence that the entanglement characteristic distance goes as the Compton wavelength of the field \cite{Zych10}, thus we should also expect the entanglement spatial distribution to hug the boundary more closely as we increase the mass $\mu$ of the field. 

To obtain information on the spatial structure of entanglement we use a technique very similar to that in \cite{Botero04}, which there was used within the context of lattice systems. Since the mode function $v_1(x)$ contains the majority of the entanglement (right now working at $t=0$), what we propose is that the function $|v_1(x)|$ gives information about the spatial structure of entanglement. The larger $|v_1(x)|$ is at a given $x$, the more entanglement is localized at that point. Operationally this proposal makes sense; if one were to try to swap this entanglement into an Unruh-deWitt type detector model then it makes sense to place the detector where $|v_1(x)|$ is largest, since this directly translates into the coupling strength between this mode and the detector. Of course there is also entanglement in the higher $v$-modes, and these would form corrections to our $|v_1|$ estimate. Seen another way, we can consider measuring the entanglement between regions by means of local projective measurements onto a pair of spatial modes \cite{Dragan13,Doukas13}. Since most of the entanglement is isolated between $v_1(x)$ and $\bar{v}_1(x)$, it is these modes that we would want to measure in order to obtain the greatest amount of entanglement.

In Fig. \ref{v1} we plot the function $|v_1(x)|$ at time $t=0$ using the parameters $r=0.5R$, $N=200$, and for three mass values $\mu$ of $0$, $10/R$, and $50/R$. As can be seen, both of the conditions discussed above are satisfied. Namely, the distribution indeed straddles the boundary between the two sides of the cavity (in this case the boundary is to the right because we are looking at the left side). Furthermore, as the mass $\mu$ of the field is increased we see that the distribution becomes more localized at the boundary, representing a decreasing correlation length. 

Note that the small vibrations that can be seen in Fig. \ref{v1} are due solely to taking a finite number $N$ of local modes. As $N$ is increased these vibrations become smaller. However the overall shape of the function does not change upon increasing $N$; a fact that further indicates that the function $|v_1(x)|$, as plotted, well represents the entanglement structure despite the mixedness problem.

All that we have done here is show the shape of the left-side mode function that contains most of the entanglement with the right side, and how much this can truly be considered a distribution of entanglement is questionable. A more thorough approach to discuss the entanglement spatial structure could be to consider the local reduced states for infinitesimally small regions and see how much these regions are entangled with the right side of the box.

\subsection{Entangled bursts of particles}

In the previous section we have looked at the form of $|v_1(x,t=0)|$ and claimed it to a good representation of the spatial distribution of entanglement. A next obvious questions is: in the case that we slam down a mirror at $t=0$, how does $|v_1(x,t)|$ evolve for $t>0$ and what significance does this have? The time evolution is simply given by Eq. (\ref{UtoV}), i.e. $v_1(x,t)$ evolves according to the Klein Gordon equation with initial conditions  given by $v_1(x,0), \dot v_1(x,0)$, as shown in Fig. \ref{v1}. As can be expected, the evolution is that of a wavepacket propagating away from the newly slammed mirror. For example in Fig. \ref{burst} we plot $|v_1(x,t)|$ at time $t=r/2$ for parameters $r=0.5R$, $N=200$, and with a field mass of $\mu=50/R$.

By construction, however, the state of this evolving mode and the correlations between it and the right-hand cavity are exactly the same as at $t=0$ (i.e. highly excited and highly entangled with right-hand mode $\bar v_1$), when these correlations could be interpreted solely as vacuum entanglement. That is, the state of the propagating wavepacket seen in Fig. \ref{burst} is highly excited, and is highly entangled with the symmetrically evolving wavepacket in the right-hand cavity. That is, we see exactly the physics we expect, namely that slamming down a mirror produces bursts of particles that propagate away from it! Similarly in the right-hand cavity the function $\bar v_1(x,t)$ represents a burst of particles propagating to the right. A detector placed within one of these cavities will then be able to measure these particles once they hit it. \footnote{One may be concerned that in Fig. \ref{burst} there appears to be an amount of acausal signaling. Of course, for a delocalized mode, it makes no sense to strictly talk about causality \cite{Brown13}. In any relevant calculation all modes would be considered and no acausal behavior would be seen.} Additionally we see that the bursts on the two sides are entangled, and that they are entangled exactly in the same manner that the vacuum was entangled prior to the introduction of the mirror! In fact, their entanglement directly results from (or rather, it simply \emph{is}) the vacuum entanglement prior to the mirror being slammed.

This emphasizes and illustrates nicely our primary message: that the real excitations created by slamming down a mirror are identical to the ``virtual" excitations attributed to the original vacuum entanglement. Furthermore, this perspective motivates an experimental approach to verifying, and perhaps even harvesting and using, vacuum entanglement. That is, if we were able to slam a mirror and measure the real particles, in such a way that we could confirm quantum correlation statistics on the two sides, then this would constitute a verification of vacuum entanglement. We discuss this further in Sect. \ref{experiment}.

\begin{figure*}[t]
	\centering
                 \includegraphics[width=\textwidth]{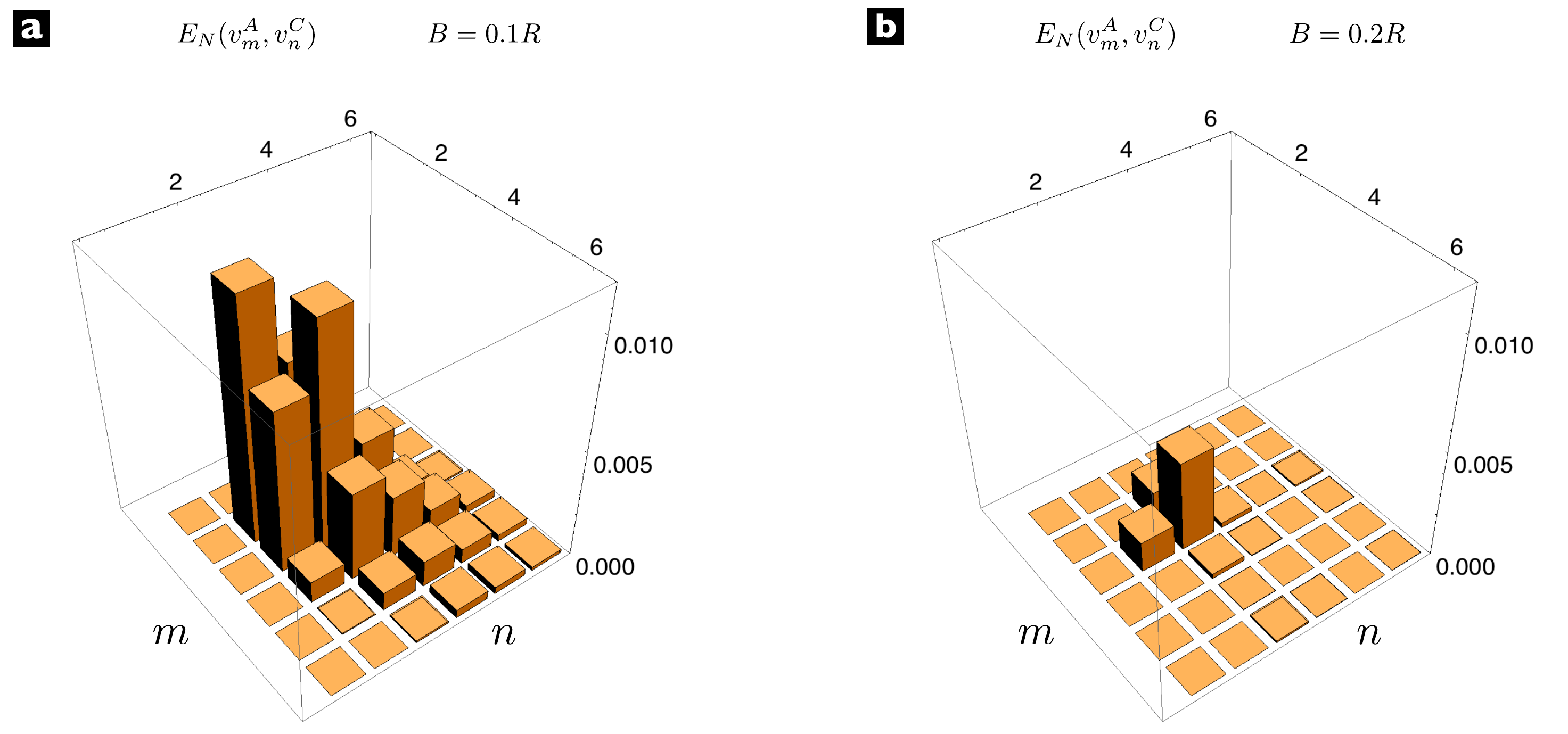}
	\caption{(Color online) Two mirror case: logarithmic negativity $E_N$ between local, diagonalizing modes $v^A_m$ and $v^C_n$ on the left and right-most sides of the cavity, respectively, in the case that the field is massless $\mu=0$. The cavity is in this case split into three regions, $\Delta_A=[0,0.5R-B/2],\  \Delta_B=[0.5R-B/2,0.5R+B/2],\ \Delta_C=[0.5R+B/2,R]$. We have taken $N=200$.  Left: Size of the middle section $B$ = 0.1R. Right: Size of the middle section $B$ = 0.2R.}
        \label{2mirrors}
\end{figure*}

The reader should know that this is an approximate picture in regards to visualizing the burst of particles, as we are just using a single delocalized mode $v_1(x,t)$. It is a good approximate picture, given that this mode contains the majority of excitations. However, in order to gain the full structure of the burst one could instead monitor the change at different times of the expectation values of local number operators attached to small (perhaps infinitesimal) regions. As the burst reaches these small regions we expect these number expectation values to jump, and they will be different from the vacuum expectation values only inside the future light cone of the spacetime point at which we slam the mirror.

\subsection{Two mirrors}

We have just stated that the entanglement between the bursts of particles produced by the slammed mirror, in the left and right-hand sides, comes from the vacuum entanglement that was previously there to begin with. One may, however, be concerned that this is simply one perspective on the situation. One may argue that what really physically occurs is that the act of slamming the mirror locally creates entangled quanta which then propagate away, rather than this entanglement having been previously present.

To debunk this view we need simply consider a slightly different scenario: that of slamming \emph{two} mirrors down simultaneously, some distance apart from each other. It is known (and we will confirm) that there is entanglement between regions of space even when they are separated. This means that when we slam two mirrors the resulting field states in the left-most and right-most cavities will be entangled, as would be measurable from the real particle statistics. In this case one cannot claim that this entanglement was simply created by the mirror, because now there is no common mirror connecting the two regions. In this case it is clear that the entanglement between the two cavities comes directly from the vacuum entanglement that was already present beforehand as no causal signal can connect them.

The mathematics of this scenario is exactly the same as before except that now we must consider splitting the cavity into three regions, as we have already discussed in Sect. \ref{stateThree}. We choose some size for the three regions (here we will take regions $A$ and $C$ to be the same size, and separated by some distance $B$). We can then take the reduced state of the left-most and right-most regions, as given by Eq. (\ref{ACstate}) and perform exactly the same entanglement analysis as we have done above. The result in short is that they are entangled. This validates our above argument since, by construction, this entanglement is present between real, stationary mode excitations after the mirrors have been introduced.

In particular, it is interesting to again perform the local, symplectic diagonalization such that we go to the local mode basis $\{v^A_m, v^C_n\}$. As discussed in Sect. \ref{SympDiag}, this procedure fails to produce a nice one-to-one entanglement structure when one's state is mixed. As we saw, the mixedness problem above only causes slight deviations from this structure. Now, however, the extra mixedness in the $AC$ system caused by tracing out $B$ \emph{really} ruins this structure. We plot in Fig. \ref{2mirrors} the mode-mode logarithmic negativity between the $v^A$ and $v^C$-modes for the cases in which the distance $B$ between the two regions is $0.1R$ and $0.2R$, where we have taken $N=200$ for each region and we use a massless field $\mu=0$. As we can see, the entanglement rapidly decays with the distance between the regions, as should be expected. We also note that in this case the \emph{higher} $v$-modes become the dominant entanglement carriers, meaning that to actually measure such entanglement one should try to change the wavepacket form that one is measuring to conform with the shape of $|v_2(x,t)|$ or $|v_3(x,t)|$ or whichever mode carries the most entanglement. It is not overly surprising that $v_1(x,t)$ becomes superseded for a large enough distance $B$ once one realizes that $v_1(x,t)$ largely contains the entanglement localized on the boundary between regions. Once there is no common boundary we therefore rapidly lose this entanglement contribution.

\section{Experimental motivations and prospects}   \label{experiment}

We would like to devote this section to discuss possible experimental platforms where to observe the phenomena here described. The primary motivation for such an experiment would be the verification of vacuum entanglement and, possibly in the future, an effective method of entanglement harvesting.

We must point out that the description of our model so far has considered an idealized theoretical scenario and has not been adapted to any particular experiment. Moreover, a first analysis shows that such an experiment would be highly challenging and  some of the requirements needed (mirror slamming times, high sensitivities...) may require considerable effort before becoming feasible. 

First of all, let us focus on the essential elements of the theoretical scheme, which should be imperatively implemented in any experiment of this sort. We require a quantum field in a cavity, which should be taken into its lowest energy state (the vacuum), and a boundary condition (here, a mirror) which will quickly appear somewhere inside the cavity and produce particles similar to the dynamical Casimir effect. For most platforms to be considered the field would be massless, as we will be dealing with electromagnetic fields. In addition, after these particles have been produced they must be detected and, if possible, their entanglement measured. 

\begin{figure}[h]
	\centering
    \includegraphics[width=0.48\textwidth]{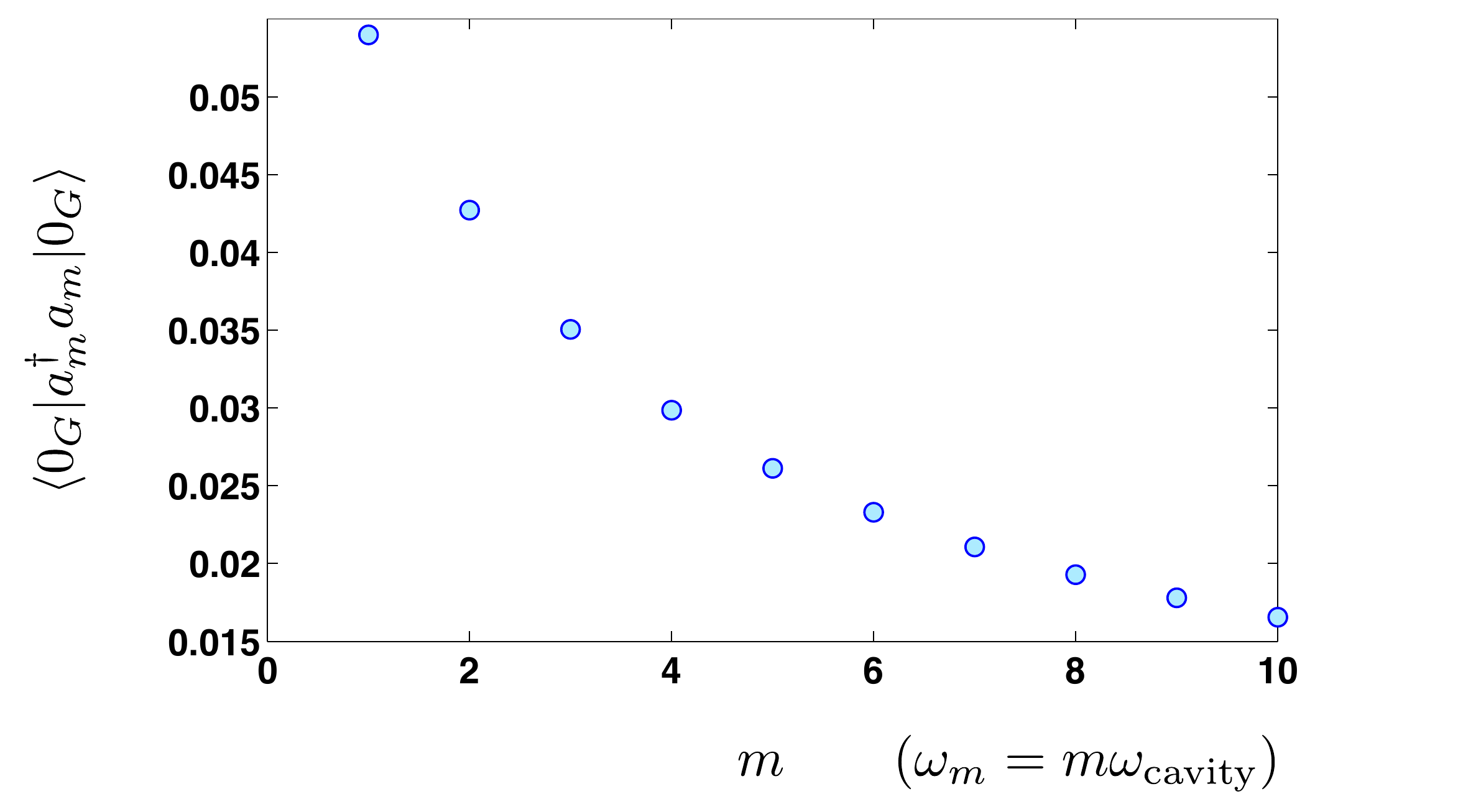}
	\caption{The number expectation value of local modes $u_m$ for the case of a massless field $\mu=0$ and a cavity split in half $r/R=0.5$.}
        \label{photonproduction}
\end{figure}

Before anything else, we should check the amount of particles created. Based on previous results \cite{Vazquez14}, Fig. \ref{photonproduction} shows the average number of local particles created after slamming the mirror, dividing the cavity in two equal sides. We can see that the maximum amount of particles creation is achieved by the first local mode, but that even this is quite small (0.052). The expected value is independent of the cavity size or the speed of the mirror, which sets only the adiabatic UV-cutoff. Any detector that aims to extract those particles would therefore be highly sensitive (and the experiment would need to be ran many times). The relative positioning of the mirror could be modified in order to improve those numbers, but that improvement is only slight and, from our point of view, not relevant enough to be discussed at this point. 

The most natural set-up for such an experiment, given the theoretical set-up, would involve the use of an optical cavity \cite{CavityQED}. In practice, however, this setup would be almost impossible to implement. In order to obtain reasonable particle production we require a slamming time that satisfies $t_\text{slam} \ll 2L/c$. For optical cavity setups this would require a slamming on the order of picoseconds, which is entirely mechanically unattainable with present technology. We conclude that this platform is unsuitable for our needs.

The more promising candidate would be Circuit Quantum Electrodynamics \cite{Blais04,You11}. Several experiments concerned with the peculiar properties of the quantum vacuum (similar to the one here discussed) have been carried on this platform. In particular the first observation of the Dynamical Casimir effect \cite{Wilson11}. The kinds of techniques used in that experiment could be very useful in a future proposal. The build up of a mirror inside the cavity, is however, a very different matter, as it implies the ``activation'' of a boundary that previously was not there. In the case of circuit QED, meandering resonators of lengths $\sim$ 20 mm have been built \cite{meandering1, meandering2} but longer lengths could be achieved, say of 100 mm. For that size a mirror slamming time of 0.7 ns may be enough to show the effects that we want.
 
Along these lines, recent work in Circuit QED \cite{Hoi11, Hoi12} has shown that a superconducting qubit coupled to a waveguide can fully reflect single photons, while it being possible to modulate the coupling to the natural mode of the cavity in the ns timescale. That could be the first candidate for the slamming mirror. However for a mirror to reflect all photons the qubit would not be enough; rather the possibility of replacing the qubit with a frequency-tunable cavity which couples to the middle of the cavity could be studied. Very fast tuning of cavities has been proven before ($\sim 3$ ns) and it is expected to be achievable in the subnanosecond regime \cite{PerDelsing}.

Finally, another experimental platform worthy of consideration would be cold atoms in optical lattices. Although we would be dealing in that case with a discrete quantum field theory (e.g. Bose-Hubbard model), the possibilities for creating ``mirror-like'' conditions by raising and lowering potential barriers using holographic techniques in the subnanosecond-picosecond regime \cite{Zupancic} might very well fit our needs.

\section{Conclusions}

We have given an answer to the question of ``what does it mean for half of an empty cavity to be full?" by considering a physical scenario in which this statement actually has operational meaning. The procedure that we considered is that of very quickly introducing one or more mirrors into a cavity scalar field prepared in its vacuum state and observing the consequences. Unsurprisingly such an action induced particle creation in the field. The key observation, however, is that these real excitations are mathematically equivalent to the local vacuum excitations related to spatial entanglement in the field. As a result, the real particles that one obtains on either side of the newly introduced mirror are entangled with each other. Furthermore we have proven that this entanglement can not simply have been created by slamming down a mirror, and rather derives directly from the previously present vacuum entanglement. We proved this by also studying the case in which two mirrors, rather than one, are slammed down simultaneously and some distance apart. In this scenario the excitations in the left-most and right-most regions created from this cavity splitting are also entangled with each other, despite there being no common mirror and no possible communication between them. This entanglement is exactly the spatial vacuum entanglement that was already present.

As part of our exposition we utilized Gaussian quantum mechanics to easily derive the reduced states and correlations of the vacuum field in different subregions of the cavity. We have used this technology to discuss the entanglement structure between regions of the cavity and the time evolution that follows upon slamming down a mirror, including directly relating the entanglement between regions with the burst of particles created by the mirror. This work provides a solid operational interpretation for vacuum entanglement and the local excitations that derive from it; these ``virtual" excitations are simply the real excitations that one gets when slamming down a mirror. In addition, this realization motivates a simple experimental proposal for the verification of vacuum entanglement in a cavity system. Indeed we discuss how the act of slamming down a mirror may represent a very effective method of harvesting the vacuum entanglement. We finished by briefly discussing some preliminary experimental prospects for the laboratory realization of this proposal.

\section*{Acknowledgements}

This work is supported by Spanish MICINN Projects FIS2011-29287 and CAM research consortium QUITEMAD S2009-ESP-1594. E. Brown acknowledges support by the Michael Smith Foreign Study Supplements Program, M. del Rey was supported by a CSIC JAE-PREDOC grant and H. Westman was supported by the CSIC JAE-DOC 2011 program. A.D. was supported by the National Science Center, Sonata BIS Grant No. 2012/07/E/ST2/01402. Furthermore we would like to thank Nicolas Menicucci and Eduardo Martin-Martinez for some fruitful discussions about this work. 
\appendix

\section{Crash course in Gaussian quantum mechanics}   \label{Gauss}

Here we very quickly review the concepts from Gaussian quantum mechanics that are required to understand the material in the main text. We do not attempt to justify or derive anything here. Everything that is presented (and much more) can be found in the active literature, for example \cite{Adesso07}.  

Let us deal with a system of $N$ continuous-variable bosonic modes (e.g. harmonic oscillators, or modes of a field). Let the annihilation and creation operators of mode $m$ be $\op a_m$ and $\op a_m^\dagger$, respectively. We can then define the internal position and momentum quadrature operators of this mode to be the canonically conjugate, Hermitian pair
\begin{align}
	\op q_m \equiv \frac{1}{\sqrt{2}}(\op a_m + \op a_m^\dagger), \;\;\;\;\; \op p_m \equiv \frac{\ii}{\sqrt{2}}(a_m^\dagger-a_m). 
\end{align}
For notational convenience we will arrange these operators into the vector $\mat{\op x}=(\op q_1, \op p_1, \op q_2, \op p_2, \cdots, \op q_N, \op p_N)^T$, with the $m$'th entry of this vector labeled $\op x_m$. In this notation the canonical commutation relations take the form
\begin{align}
	[\op x_m, \op x_n]=\ii \Omega_{mn},
\end{align}
where $\Omega_{mn}$ are the entries of a matrix called the \emph{symplectic form}, which is given by
\begin{align}
	\mat \Omega = \bigoplus_{m=1}^N
	\begin{pmatrix}
		0 & 1 \\
		-1 & 0
	\end{pmatrix}.
\end{align}

The state $\op \rho$ of our system is said to be \emph{Gaussian} if its corresponding Wigner function is Gaussian over phase space. Equivalently, the state is fully characterized by the first and second moments, $\braket{\op x_m}=\text{Tr}(\op \rho \op x_m)$ and $\braket{\op x_m \op x_n}=\text{Tr}(\op \rho \op x_m \op x_n)$. In this work we only need to consider states that have zero-mean (i.e. zero first moments). In this case only the second moments are required. Thus, rather than using the density operator to characterize the state we instead use the $2N \times 2N$ \emph{covariance matrix} $\mat \sigma$, the entries of which are defined to be
\begin{align}
	\sigma_{m n}=\braket{\op x_m \op x_n+\op x_n \op x_m}.
\end{align}
We use this matrix to fully characterize the state.

We can decompose the covariance matrix in $2\times 2$ blocks:
\begin{align}
	\mat \sigma= \begin{pmatrix}
		\mat \sigma_{11} & \mat \sigma_{12} & \cdots \\
		\mat \sigma_{21}^T & \mat \sigma_{22} & \cdots \\
		\vdots & \vdots & \ddots 
	\end{pmatrix}.
\end{align}
Here the matrix $\mat \sigma_{mm}$ is in fact the covariance matrix (i.e. reduced state) of mode $m$. Similarly, $\mat \sigma_{mn}$ contains information about the correlations (e.g. entanglement) between modes $m$ and $n$, which are completely uncorrelated (i.e. in a product state) iff $\mat \sigma_{mn}=\mat 0$. Taking a partial trace within the covariance matrix formalism is entirely trivial; for example the reduced state of the first two modes is simply the upper-left $4\times 4$ block of $\mat \sigma$.

An important example of a Gaussian state is the ground (vacuum) state of the free Hamiltonian $\op H =\sum_{m=1}^N \omega_m \op a_m^\dagger \op a_m$. For this state the covariance matrix is straightforwardly seen to be given by the identity: $\mat \sigma_\text{vac}= \mat I$.

In general, unitary transformations $\op U$ in the Hilbert space that are generated by quadratic Hamiltonians preserve Gaussianity. Such transformations are represented by a \emph{symplectic} transformation $\mat S$ in the phase space. Namely, such a quadratically generated $\op U$ transforms the elements $\op x_m$ to a new set of quadratures $\mat{\op x}'=\op U \mat{\op x} \op U^\dagger$ such that the new quadratures are a linear combination of the old: $\mat {\op x'}=\mat S \mat{\op x}$, where in order to preserve the canonical commutation relations the matrix $\mat S$ must be symplectic,
\begin{align}
	\mat S \mat \Omega \mat S^T = \mat S^T \mat \Omega \mat S = \mat \Omega.
\end{align}
In addition, a matrix must be square in order to be considered symplectic, meaning that it transforms $N$ modes to $N$ modes. It is easily seen that on the level of the covariance matrix this transformation takes the form
\begin{align}
	\mat{\sigma} \rightarrow \mat{\sigma'}=\mat S \mat \sigma \mat S^T.
\end{align}

An important characterization of a given Gaussian state is its \emph{symplectic spectrum}. Every $N$-mode Gaussian state $\mat \sigma$ has $N$ \emph{symplectic eigenvalues} $\{\nu_m\}$, which are invariant under symplectic transformations. The covariance matrix is \emph{symplectically diagonalizable}, meaning that there exists a symplectic matrix $\mat S$ which brings the state to a diagonal form given by $\mat S \mat \sigma \mat S^T=\mat D=\text{diag} (\nu_1,\nu_1,\nu_2,\nu_2, \cdots, \nu_N, \nu_N)$. This diagonalized form is also known as the \emph{Williamson normal form} of the state. We note, for example, that the vacuum state $\mat \sigma_\text{vac}=\mat I$ is already in its Williamson normal form, and that all of its symplectic eigenvalues are equal to unity.
Note that the symplectic eigenvalues of $\mat \sigma$ are \emph{not} the same as its regular eigenvalues.

The symplectic eigenvalues of a state must always be larger than or equal to unity: $\nu_m \geq 1 \; \forall \; m$. This is simply a statement of the uncertainty principle, which is saturated iff all symplectic eigenvalues are equal to unity. The symplectic spectrum also specifies the mixedness of a Gaussian state: such a state is pure iff all symplectic eigenvalues are equal to unity. That is, a pure Gaussian state saturates the uncertainty principle. Any uncertainty in the state beyond this must be caused by classical uncertainty, i.e. mixedness. An informationally rigorous measure of mixedness, the von Neumann entropy $S(\mat \sigma)$ of the state, can be computed from the symplectic eigenvalues via
\begin{align} \label{entropy}
	S(\mat \sigma)=\sum_{m=1}^N f(\nu_m),
\end{align}
where
\begin{align}
	f(x)=\frac{x+1}{2}\log\left(\frac{x+1}{2}\right)-\frac{x-1}{2}\log\left(\frac{x-1}{2}\right).
\end{align}
The entropy is zero for a pure state, when $\nu_m=1$ for all $m$.

The easiest way to compute the symplectic eigenvalues of a state (if one does not care about the diagonalizing transformation) is to compute the regular eigenvalues of the matrix $\ii \mat \Omega \mat \sigma$, which come in pairs of $\{\pm \nu_m\}$. There are situations, however, in which one would also like to compute the diagonalizing symplectic transformation itself. The method of doing this is provided in Appendix \ref{diagonalization}. Of particular importance for us is the joint, local diagonalization of a bipartite, pure state. Imagine that we split our set of modes into two groups, $A$ and $B$. The joint state can then be decomposed as
\begin{align}
	\mat \sigma = \begin{pmatrix}
		\mat \sigma_A & \mat \gamma \\
		\mat \gamma^T & \mat \sigma_B
	\end{pmatrix},
\end{align}
where $\mat \sigma_A$ and $\mat \sigma_B$ are the reduced states for groups $A$ and $B$, respectively, and $\mat \gamma$ contains the correlations between the two groups. Let us assume that the global state is pure. That is, we assume that the symplectic eigenvalues $\nu_m$ of $\mat \sigma$ are all equal to unity. This does not mean, however, that the symplectic eigenvalues of $\mat \sigma_A$ and $\mat \sigma_B$ are all equal to unity; indeed they will not be if the bipartitions are entangled. Let us label the ``local" symplectic eigenvalues of these reduced states as $\nu_m^{(A)}$ and $\nu_m^{(B)}$. Because $\mat \sigma$ is pure these two spectrums will in fact be equivalent (with the larger of the two systems having extra symplectic eigenvalues equal to unity); this is equivalent to the fact that the standard local spectrums of reduced density operators in a pure bipartition are equal. Let $\mat S_A$ be the local symplectic transformation that diagonalizes $\mat \sigma_A$, and similarly we have $\mat S_B$. Let us then apply these local transformations to our state by acting on $\mat \sigma$ with the joint transformation $\mat S_A \oplus \mat S_B$:
\begin{align}
	(\mat S_A \oplus \mat S_B)\mat \sigma (\mat S_A \oplus \mat S_B)^T=\begin{pmatrix}
		\mat D_A & \mat \gamma_D    \\
		\mat \gamma_D^T & \mat D_B
	\end{pmatrix}.
\end{align}
The reduced states have now been put into their Williamson normal forms. Because this is a purely local operation the entanglement between the two sides has not been modified. Importantly, \emph{if} the global state is pure then this transformation produces a correlation matrix $\mat \gamma_D$ that is diagonal as well \cite{Adesso07}. This is analogous to the Hilbert space Schmidt decomposition of a pure, bipartite state. In the literature on Gaussian quantum mechanics such a covariance matrix is said to be in \emph{standard form}. The fact that $\mat \gamma_D$ is diagonal means that in this locally transformed basis we obtain a product of pure, two-mode states. That is, each pair is uncorrelated with any others. Generally each such pair of modes will be entangled (in particular, they will be in a two-mode squeezed state). Performing this local symplectic diagonalization is therefore a method of isolating the entanglement between $A$ and $B$ into simple pairs of modes (rather than the entanglement between a given mode in $A$ and the rest of the system being spread across multiple modes in both $A$ and $B$).

In the case that $\mat \sigma$ is mixed we unfortunately cannot perform the same feat. We can, of course, still locally diagonalize the reduced systems. This removes any mode-mode correlation within $A$ and $B$ themselves. However in this case the resulting correlation matrix $\mat \gamma_D$ will not generally be diagonal, meaning that we can still have a given mode in $A$ being correlated with multiple modes in $B$, and vice versa.

Lastly, we wish to have a measure of entanglement in Gaussian states. In the case of a globally pure state the entanglement across a bipartition is simply the entropy, Eq. (\ref{entropy}), of either of the two reduced states. In the case in which the state is globally mixed, on the other hand, one can use the logarithmic negativity $E_N$ \cite{Plenio05,Adesso07}. For bipartite Gaussian states a non-zero value of $E_N$ is a necessary and sufficient condition for non-separability \cite{Simon00}. For a two-mode Gaussian state with covariance matrix
\begin{align}
	\mat \sigma_\text{two mode}=\begin{pmatrix}
		\mat \sigma_{11} & \mat \sigma_{12} \\
		\mat \sigma_{21}^T & \mat \sigma_{22}
	\end{pmatrix}
\end{align}
the logarithmic negativity between the modes is given by
\begin{align}   \label{logneg}
	E_N=\max(0,-\log z),
\end{align}
where
\begin{align}
	2z^2=\Delta-\sqrt{\Delta^2-4 \det \mat \sigma_\text{two mode}},
\end{align}
and where $\Delta=\det \mat \sigma_{11}+\det \mat \sigma_{22}-2\det \mat \sigma_{12}$.

\section{Symplectic diagonalization}  \label{diagonalization}

Here we describe the method of symplectically diagonalizing a covariance matrix, i.e. putting into its Williamson normal form. To do this it is easier to work in a re-ordered phase space basis in which the $q$'s are packaged together and similarly for the $p$'s: $\op{\mat x}=(\op q_1, \op q_2, \cdots, \op p_1, \op p_2, \cdots)$. In this basis the reduced covariance matrix of Eq. (\ref{covBlock}), for example, takes a block form
\begin{align} \label{covmat2}
	\mat \sigma = \begin{pmatrix}
 		\mat \sigma^{(Q)} & \mat 0 \\
		\mat 0 & \mat \sigma^{(P)}
	\end{pmatrix},
\end{align}
where the entries of these blocks, $\sigma^{(Q)}_{m n}$ and $\sigma^{(P)}_{mn}$, are given by the upper left and lower right entries of $\mat \sigma_{mn}$ in Eq. (\ref{covmatR}), respectively. The off-diagonal blocks of Eq. (\ref{covmat2}) are zero due to the fact that the Bogoliubov transformation to the local basis is purely real. This circumstance in fact makes it considerably easier to symplectically diagonalize $\mat \sigma$, and here we will only cover this case.

Note that in the new basis ordering the symplectic form is given by
\begin{align}
	\mat \Omega=\begin{pmatrix}
	\mat 0 & \mat I \\
	-\mat I & \mat 0
	\end{pmatrix}.
\end{align}
Also in this basis the Williamson normal (symplectically diagonalized) form of a covariance matrix is given by $\mat D =\mat \nu \oplus \mat \nu$, where $\mat \nu = \text{diag}(\nu_1,\nu_2, \cdots)$ contains the symplectic eigenvalues.

We would like to find the symplectic transformation $\mat{\tilde S}$ that achieves this transformation. Specifically we will let $\mat{\tilde S}^T  (\mat \nu \oplus \mat \nu) \mat{\tilde S}=\mat \sigma$. To this end, we will make an Ansatz and then prove that it is the correct choice. Let us define a matrix $\mat A \equiv \sqrt{\mat \sigma^{(Q)}}\sqrt{\mat \sigma^{(P)}}$. We claim that the symplectic eigenvalues $\{\nu_m\}$ of $\mat \sigma$ are given by the singular values of $\mat A$. That is, there are orthogonal matrices $\mat O_1$ and $\mat O_2$ such that
\begin{align} \label{svd}
	\mat A = \mat O_1^T \mat \nu \mat O_2.
\end{align}
Let us take these and form another orthogonal matrix given by their direct sum $\mat O \equiv \mat O_1 \oplus \mat O_2$. We now claim that the symplectic matrix $\mat{\tilde S}$ that diagonalizes $\mat \sigma$ is given by
\begin{align}
	\mat{\tilde S}=(\mat \nu \oplus \mat \nu)^{-1/2} \mat O \mat \sigma^{1/2}.
\end{align}
Clearly from this definition it is true that $\mat{\tilde S}^T  (\mat \nu \oplus \mat \nu) \mat{\tilde S}=\mat \sigma$, since $\mat O$ is orthogonal. However, is it symplectic: $\mat{\tilde S}\mat \Omega \mat{\tilde S}^T=\mat \Omega$? By expanding the left-hand side of this equation it is straightforward to see that the transformation will be symplectic iff $\mat O_1 \mat A \mat O_2^T=\mat \nu$, which is equivalent to Eq. (\ref{svd}).

Thus, finding the symplectic diagonalization is equivalent to finding the singular value decomposition of the matrix $\mat A$, which is easily done numerically. Note that to go from the matrix $\mat \sigma$ to $\mat \nu \oplus \mat \nu$ in the sense of $\mat S_D \mat \sigma \mat S_D^T=\mat \nu \oplus \mat \nu$, the correct transformation will be $\mat S_D = (\mat{\tilde S}^T)^{-1}=(\mat \nu \oplus \mat \nu)^{1/2} \mat O \mat \sigma^{-1/2}$.

\bibliography{halfbox_ver6}

\begin{thebibliography}{37}%
\makeatletter
\providecommand \@ifxundefined [1]{%
 \@ifx{#1\undefined}
}%
\providecommand \@ifnum [1]{%
 \ifnum #1\expandafter \@firstoftwo
 \else \expandafter \@secondoftwo
 \fi
}%
\providecommand \@ifx [1]{%
 \ifx #1\expandafter \@firstoftwo
 \else \expandafter \@secondoftwo
 \fi
}%
\providecommand \natexlab [1]{#1}%
\providecommand \enquote  [1]{``#1''}%
\providecommand \bibnamefont  [1]{#1}%
\providecommand \bibfnamefont [1]{#1}%
\providecommand \citenamefont [1]{#1}%
\providecommand \href@noop [0]{\@secondoftwo}%
\providecommand \href [0]{\begingroup \@sanitize@url \@href}%
\providecommand \@href[1]{\@@startlink{#1}\@@href}%
\providecommand \@@href[1]{\endgroup#1\@@endlink}%
\providecommand \@sanitize@url [0]{\catcode `\\12\catcode `\$12\catcode
  `\&12\catcode `\#12\catcode `\^12\catcode `\_12\catcode `\%12\relax}%
\providecommand \@@startlink[1]{}%
\providecommand \@@endlink[0]{}%
\providecommand \url  [0]{\begingroup\@sanitize@url \@url }%
\providecommand \@url [1]{\endgroup\@href {#1}{\urlprefix }}%
\providecommand \urlprefix  [0]{URL }%
\providecommand \Eprint [0]{\href }%
\providecommand \doibase [0]{http://dx.doi.org/}%
\providecommand \selectlanguage [0]{\@gobble}%
\providecommand \bibinfo  [0]{\@secondoftwo}%
\providecommand \bibfield  [0]{\@secondoftwo}%
\providecommand \translation [1]{[#1]}%
\providecommand \BibitemOpen [0]{}%
\providecommand \bibitemStop [0]{}%
\providecommand \bibitemNoStop [0]{.\EOS\space}%
\providecommand \EOS [0]{\spacefactor3000\relax}%
\providecommand \BibitemShut  [1]{\csname bibitem#1\endcsname}%
\let\auto@bib@innerbib\@empty
\bibitem [{\citenamefont {Summers}\ and\ \citenamefont
  {Werner}(1987)}]{Summers87}%
  \BibitemOpen
  \bibfield  {author} {\bibinfo {author} {\bibfnamefont {S.~J.}\ \bibnamefont
  {Summers}}\ and\ \bibinfo {author} {\bibfnamefont {R.}~\bibnamefont
  {Werner}},\ }\href@noop {} {\bibfield  {journal} {\bibinfo  {journal}
  {Commun. Math. Phys}\ }\textbf {\bibinfo {volume} {110}},\ \bibinfo {pages}
  {247} (\bibinfo {year} {1987})}\BibitemShut {NoStop}%
\bibitem [{\citenamefont {Srednicki}(1993)}]{Srednicki93}%
  \BibitemOpen
  \bibfield  {author} {\bibinfo {author} {\bibfnamefont {M.}~\bibnamefont
  {Srednicki}},\ }\href@noop {} {\bibfield  {journal} {\bibinfo  {journal}
  {Phys. Rev. Lett}\ }\textbf {\bibinfo {volume} {71}},\ \bibinfo {pages} {666}
  (\bibinfo {year} {1993})}\BibitemShut {NoStop}%
\bibitem [{\citenamefont {Casini}\ and\ \citenamefont
  {Huerta}(2009)}]{Casini09}%
  \BibitemOpen
  \bibfield  {author} {\bibinfo {author} {\bibfnamefont {H.}~\bibnamefont
  {Casini}}\ and\ \bibinfo {author} {\bibfnamefont {M.}~\bibnamefont
  {Huerta}},\ }\href@noop {} {\bibfield  {journal} {\bibinfo  {journal} {J.
  Phys. A: Math. Theor.}\ }\textbf {\bibinfo {volume} {42}},\ \bibinfo {pages}
  {504007} (\bibinfo {year} {2009})}\BibitemShut {NoStop}%
\bibitem [{\citenamefont {Reznik}\ \emph
  {et~al.}(2005{\natexlab{a}})\citenamefont {Reznik}, \citenamefont {Retzker},\
  and\ \citenamefont {Silman}}]{Reznik05}%
  \BibitemOpen
  \bibfield  {author} {\bibinfo {author} {\bibfnamefont {B.}~\bibnamefont
  {Reznik}}, \bibinfo {author} {\bibfnamefont {A.}~\bibnamefont {Retzker}}, \
  and\ \bibinfo {author} {\bibfnamefont {J.}~\bibnamefont {Silman}},\
  }\href@noop {} {\bibfield  {journal} {\bibinfo  {journal} {Phys. Rev. A}\
  }\textbf {\bibinfo {volume} {71}},\ \bibinfo {pages} {042104} (\bibinfo
  {year} {2005}{\natexlab{a}})}\BibitemShut {NoStop}%
\bibitem [{\citenamefont {Braun}(2005)}]{Braun05}%
  \BibitemOpen
  \bibfield  {author} {\bibinfo {author} {\bibfnamefont {D.}~\bibnamefont
  {Braun}},\ }\href@noop {} {\bibfield  {journal} {\bibinfo  {journal} {Phys.
  Rev. A}\ }\textbf {\bibinfo {volume} {72}},\ \bibinfo {pages} {062324}
  (\bibinfo {year} {2005})}\BibitemShut {NoStop}%
\bibitem [{\citenamefont {Leon}\ and\ \citenamefont {Sabin}(2009)}]{Leon09}%
  \BibitemOpen
  \bibfield  {author} {\bibinfo {author} {\bibfnamefont {J.}~\bibnamefont
  {Leon}}\ and\ \bibinfo {author} {\bibfnamefont {C.}~\bibnamefont {Sabin}},\
  }\href@noop {} {\bibfield  {journal} {\bibinfo  {journal} {Phys. Rev. A}\
  }\textbf {\bibinfo {volume} {79}},\ \bibinfo {pages} {012304} (\bibinfo
  {year} {2009})}\BibitemShut {NoStop}%
\bibitem [{\citenamefont {VerSteeg}\ and\ \citenamefont
  {Menicucci}(2009)}]{Steeg09}%
  \BibitemOpen
  \bibfield  {author} {\bibinfo {author} {\bibfnamefont {G.}~\bibnamefont
  {VerSteeg}}\ and\ \bibinfo {author} {\bibfnamefont {N.~C.}\ \bibnamefont
  {Menicucci}},\ }\href@noop {} {\bibfield  {journal} {\bibinfo  {journal}
  {Phys. Rev. D}\ }\textbf {\bibinfo {volume} {79}},\ \bibinfo {pages} {044027}
  (\bibinfo {year} {2009})}\BibitemShut {NoStop}%
\bibitem [{\citenamefont {Brown}(2013)}]{Brown13-3}%
  \BibitemOpen
  \bibfield  {author} {\bibinfo {author} {\bibfnamefont {E.~G.}\ \bibnamefont
  {Brown}},\ }\href@noop {} {\bibfield  {journal} {\bibinfo  {journal} {Phys.
  Rev. A}\ }\textbf {\bibinfo {volume} {88}},\ \bibinfo {pages} {062336}
  (\bibinfo {year} {2013})}\BibitemShut {NoStop}%
\bibitem [{\citenamefont {Cramer}\ \emph {et~al.}(2006)\citenamefont {Cramer},
  \citenamefont {Eisert}, \citenamefont {Plenio},\ and\ \citenamefont
  {Dreibig}}]{Cramer06}%
  \BibitemOpen
  \bibfield  {author} {\bibinfo {author} {\bibfnamefont {M.}~\bibnamefont
  {Cramer}}, \bibinfo {author} {\bibfnamefont {J.}~\bibnamefont {Eisert}},
  \bibinfo {author} {\bibfnamefont {M.~B.}\ \bibnamefont {Plenio}}, \ and\
  \bibinfo {author} {\bibfnamefont {J.}~\bibnamefont {Dreibig}},\ }\href@noop
  {} {\bibfield  {journal} {\bibinfo  {journal} {Phys. Rev. A}\ }\textbf
  {\bibinfo {volume} {73}},\ \bibinfo {pages} {012309} (\bibinfo {year}
  {2006})}\BibitemShut {NoStop}%
\bibitem [{\citenamefont {Amico}\ \emph {et~al.}(2008)\citenamefont {Amico},
  \citenamefont {Fazio}, \citenamefont {Osterloh},\ and\ \citenamefont
  {Vedral}}]{Amico08}%
  \BibitemOpen
  \bibfield  {author} {\bibinfo {author} {\bibfnamefont {L.}~\bibnamefont
  {Amico}}, \bibinfo {author} {\bibfnamefont {R.}~\bibnamefont {Fazio}},
  \bibinfo {author} {\bibfnamefont {A.}~\bibnamefont {Osterloh}}, \ and\
  \bibinfo {author} {\bibfnamefont {V.}~\bibnamefont {Vedral}},\ }\href@noop {}
  {\bibfield  {journal} {\bibinfo  {journal} {Rev. Mod. Phys.}\ }\textbf
  {\bibinfo {volume} {80}},\ \bibinfo {pages} {517} (\bibinfo {year}
  {2008})}\BibitemShut {NoStop}%
\bibitem [{\citenamefont {Retzker}\ \emph {et~al.}(2005)\citenamefont
  {Retzker}, \citenamefont {Circa},\ and\ \citenamefont {Reznik}}]{Retzker05}%
  \BibitemOpen
  \bibfield  {author} {\bibinfo {author} {\bibfnamefont {A.}~\bibnamefont
  {Retzker}}, \bibinfo {author} {\bibfnamefont {J.~I.}\ \bibnamefont {Circa}},
  \ and\ \bibinfo {author} {\bibfnamefont {B.}~\bibnamefont {Reznik}},\
  }\href@noop {} {\bibfield  {journal} {\bibinfo  {journal} {Phys. Rev. Lett}\
  }\textbf {\bibinfo {volume} {94}},\ \bibinfo {pages} {050504} (\bibinfo
  {year} {2005})}\BibitemShut {NoStop}%
\bibitem [{\citenamefont {Birrel}\ and\ \citenamefont
  {Davies}(1984)}]{BirrelDavies84}%
  \BibitemOpen
  \bibfield  {author} {\bibinfo {author} {\bibfnamefont {N.~D.}\ \bibnamefont
  {Birrel}}\ and\ \bibinfo {author} {\bibfnamefont {P.~C.~W.}\ \bibnamefont
  {Davies}},\ }\href@noop {} {\emph {\bibinfo {title} {Quantum Fields in Curved
  Space}}}\ (\bibinfo  {publisher} {Cambridge University Press},\ \bibinfo
  {year} {1984})\BibitemShut {NoStop}%
\bibitem [{\citenamefont {Moore}(1970)}]{Moore70}%
  \BibitemOpen
  \bibfield  {author} {\bibinfo {author} {\bibfnamefont {G.}~\bibnamefont
  {Moore}},\ }\href@noop {} {\bibfield  {journal} {\bibinfo  {journal} {J.
  Math. Phys.}\ }\textbf {\bibinfo {volume} {11}},\ \bibinfo {pages} {2679}
  (\bibinfo {year} {1970})}\BibitemShut {NoStop}%
\bibitem [{\citenamefont {Wilson}\ \emph {et~al.}(2011)\citenamefont {Wilson},
  \citenamefont {Johansson}, \citenamefont {Pourkabirian}, \citenamefont
  {Simoen}, \citenamefont {Johansson}, \citenamefont {Duty}, \citenamefont
  {Nori},\ and\ \citenamefont {Delsing}}]{Wilson11}%
  \BibitemOpen
  \bibfield  {author} {\bibinfo {author} {\bibfnamefont {C.~M.}\ \bibnamefont
  {Wilson}}, \bibinfo {author} {\bibfnamefont {G.}~\bibnamefont {Johansson}},
  \bibinfo {author} {\bibfnamefont {A.}~\bibnamefont {Pourkabirian}}, \bibinfo
  {author} {\bibfnamefont {M.}~\bibnamefont {Simoen}}, \bibinfo {author}
  {\bibfnamefont {J.~R.}\ \bibnamefont {Johansson}}, \bibinfo {author}
  {\bibfnamefont {T.}~\bibnamefont {Duty}}, \bibinfo {author} {\bibfnamefont
  {F.}~\bibnamefont {Nori}}, \ and\ \bibinfo {author} {\bibfnamefont
  {P.}~\bibnamefont {Delsing}},\ }\href {http://dx.doi.org/10.1038/nature10561}
  {\bibfield  {journal} {\bibinfo  {journal} {Nature}\ }\textbf {\bibinfo
  {volume} {479}},\ \bibinfo {pages} {376} (\bibinfo {year}
  {2011})}\BibitemShut {NoStop}%
\bibitem [{\citenamefont {Rodriguez-Vazquez}\ \emph {et~al.}(2014)\citenamefont
  {Rodriguez-Vazquez}, \citenamefont {del Rey}, \citenamefont {Westman},\ and\
  \citenamefont {Leon}}]{Vazquez14}%
  \BibitemOpen
  \bibfield  {author} {\bibinfo {author} {\bibfnamefont {M.}~\bibnamefont
  {Rodriguez-Vazquez}}, \bibinfo {author} {\bibfnamefont {M.}~\bibnamefont {del
  Rey}}, \bibinfo {author} {\bibfnamefont {H.}~\bibnamefont {Westman}}, \ and\
  \bibinfo {author} {\bibfnamefont {J.}~\bibnamefont {Leon}},\ }\href@noop {}
  {\bibfield  {journal} {\bibinfo  {journal} {Preprint - arXiv:1403.0073
  (Annals of Physics - In Press)}\ } (\bibinfo {year} {2014})}\BibitemShut
  {NoStop}%
\bibitem [{\citenamefont {Adesso}\ and\ \citenamefont
  {Illuminati}(2007)}]{Adesso07}%
  \BibitemOpen
  \bibfield  {author} {\bibinfo {author} {\bibfnamefont {G.}~\bibnamefont
  {Adesso}}\ and\ \bibinfo {author} {\bibfnamefont {F.}~\bibnamefont
  {Illuminati}},\ }\href@noop {} {\bibfield  {journal} {\bibinfo  {journal} {J.
  Phys. A: Math. Theor}\ }\textbf {\bibinfo {volume} {40}},\ \bibinfo {pages}
  {7821} (\bibinfo {year} {2007})}\BibitemShut {NoStop}%
\bibitem [{\citenamefont {Botero}\ and\ \citenamefont
  {Reznik}(2004)}]{Botero04}%
  \BibitemOpen
  \bibfield  {author} {\bibinfo {author} {\bibfnamefont {A.}~\bibnamefont
  {Botero}}\ and\ \bibinfo {author} {\bibfnamefont {B.}~\bibnamefont
  {Reznik}},\ }\href@noop {} {\bibfield  {journal} {\bibinfo  {journal} {Phys.
  Rev. A}\ }\textbf {\bibinfo {volume} {70}},\ \bibinfo {pages} {052329}
  (\bibinfo {year} {2004})}\BibitemShut {NoStop}%
\bibitem [{\citenamefont {Reznik}\ \emph
  {et~al.}(2005{\natexlab{b}})\citenamefont {Reznik}, \citenamefont {Retzker},\
  and\ \citenamefont {Silman}}]{ReznikRetzker05}%
  \BibitemOpen
  \bibfield  {author} {\bibinfo {author} {\bibfnamefont {B.}~\bibnamefont
  {Reznik}}, \bibinfo {author} {\bibfnamefont {A.}~\bibnamefont {Retzker}}, \
  and\ \bibinfo {author} {\bibfnamefont {J.}~\bibnamefont {Silman}},\ }\href
  {\doibase 10.1103/PhysRevA.71.042104} {\bibfield  {journal} {\bibinfo
  {journal} {Phys. Rev. A}\ }\textbf {\bibinfo {volume} {71}},\ \bibinfo
  {pages} {042104} (\bibinfo {year} {2005}{\natexlab{b}})}\BibitemShut
  {NoStop}%
\bibitem [{\citenamefont {Mart\'in-Mart\'inez}\ \emph
  {et~al.}(2013)\citenamefont {Mart\'in-Mart\'inez}, \citenamefont {Brown},
  \citenamefont {Donnelly},\ and\ \citenamefont {Kempf}}]{EduEric13}%
  \BibitemOpen
  \bibfield  {author} {\bibinfo {author} {\bibfnamefont {E.}~\bibnamefont
  {Mart\'in-Mart\'inez}}, \bibinfo {author} {\bibfnamefont {E.~G.}\
  \bibnamefont {Brown}}, \bibinfo {author} {\bibfnamefont {W.}~\bibnamefont
  {Donnelly}}, \ and\ \bibinfo {author} {\bibfnamefont {A.}~\bibnamefont
  {Kempf}},\ }\href {\doibase 10.1103/PhysRevA.88.052310} {\bibfield  {journal}
  {\bibinfo  {journal} {Phys. Rev. A}\ }\textbf {\bibinfo {volume} {88}},\
  \bibinfo {pages} {052310} (\bibinfo {year} {2013})}\BibitemShut {NoStop}%
\bibitem [{\citenamefont {Unruh}(1976)}]{Unruh}%
  \BibitemOpen
  \bibfield  {author} {\bibinfo {author} {\bibfnamefont {W.~G.}\ \bibnamefont
  {Unruh}},\ }\href {\doibase 10.1103/PhysRevD.14.870} {\bibfield  {journal}
  {\bibinfo  {journal} {Phys. Rev. D}\ }\textbf {\bibinfo {volume} {14}},\
  \bibinfo {pages} {870} (\bibinfo {year} {1976})}\BibitemShut {NoStop}%
\bibitem [{\citenamefont {Brown}\ \emph {et~al.}(2013)\citenamefont {Brown},
  \citenamefont {Martin-Martinez}, \citenamefont {Menicucci},\ and\
  \citenamefont {Mann}}]{Brown13}%
  \BibitemOpen
  \bibfield  {author} {\bibinfo {author} {\bibfnamefont {E.~G.}\ \bibnamefont
  {Brown}}, \bibinfo {author} {\bibfnamefont {E.}~\bibnamefont
  {Martin-Martinez}}, \bibinfo {author} {\bibfnamefont {N.~C.}\ \bibnamefont
  {Menicucci}}, \ and\ \bibinfo {author} {\bibfnamefont {R.~B.}\ \bibnamefont
  {Mann}},\ }\href@noop {} {\bibfield  {journal} {\bibinfo  {journal} {Phys.
  Rev. D}\ }\textbf {\bibinfo {volume} {87}},\ \bibinfo {pages} {084062}
  (\bibinfo {year} {2013})}\BibitemShut {NoStop}%
\bibitem [{\citenamefont {del Rey}\ \emph {et~al.}(2012)\citenamefont {del
  Rey}, \citenamefont {Sabin},\ and\ \citenamefont {Leon}}]{delRey12}%
  \BibitemOpen
  \bibfield  {author} {\bibinfo {author} {\bibfnamefont {M.}~\bibnamefont {del
  Rey}}, \bibinfo {author} {\bibfnamefont {C.}~\bibnamefont {Sabin}}, \ and\
  \bibinfo {author} {\bibfnamefont {J.}~\bibnamefont {Leon}},\ }\href {\doibase
  10.1103/PhysRevA.85.045802} {\bibfield  {journal} {\bibinfo  {journal} {Phys.
  Rev. A}\ }\textbf {\bibinfo {volume} {85}},\ \bibinfo {pages} {045802}
  (\bibinfo {year} {2012})}\BibitemShut {NoStop}%
\bibitem [{\citenamefont {Bogoliubov}(1947)}]{Bogos}%
  \BibitemOpen
  \bibfield  {author} {\bibinfo {author} {\bibfnamefont {N.~N.}\ \bibnamefont
  {Bogoliubov}},\ }\href@noop {} {\bibfield  {journal} {\bibinfo  {journal} {J.
  Phys. (USSR)}\ }\textbf {\bibinfo {volume} {11}},\ \bibinfo {pages} {23}
  (\bibinfo {year} {1947})}\BibitemShut {NoStop}%
\bibitem [{\citenamefont {Plenio}(2005)}]{Plenio05}%
  \BibitemOpen
  \bibfield  {author} {\bibinfo {author} {\bibfnamefont {M.~B.}\ \bibnamefont
  {Plenio}},\ }\href@noop {} {\bibfield  {journal} {\bibinfo  {journal} {Phys.
  Rev. Lett}\ }\textbf {\bibinfo {volume} {95}},\ \bibinfo {pages} {090503}
  (\bibinfo {year} {2005})}\BibitemShut {NoStop}%
\bibitem [{\citenamefont {Zych}\ \emph {et~al.}(2010)\citenamefont {Zych},
  \citenamefont {Costa}, \citenamefont {Kofler},\ and\ \citenamefont
  {Brukner}}]{Zych10}%
  \BibitemOpen
  \bibfield  {author} {\bibinfo {author} {\bibfnamefont {M.}~\bibnamefont
  {Zych}}, \bibinfo {author} {\bibfnamefont {F.}~\bibnamefont {Costa}},
  \bibinfo {author} {\bibfnamefont {J.}~\bibnamefont {Kofler}}, \ and\ \bibinfo
  {author} {\bibfnamefont {C.}~\bibnamefont {Brukner}},\ }\href@noop {}
  {\bibfield  {journal} {\bibinfo  {journal} {Phys. Rev. D}\ }\textbf {\bibinfo
  {volume} {81}},\ \bibinfo {pages} {125019} (\bibinfo {year}
  {2010})}\BibitemShut {NoStop}%
\bibitem [{\citenamefont {Dragan}\ \emph {et~al.}(2013)\citenamefont {Dragan},
  \citenamefont {Doukas}, \citenamefont {Mart\'in-Mart\'inez},\ and\
  \citenamefont {Bruschi}}]{Dragan13}%
  \BibitemOpen
  \bibfield  {author} {\bibinfo {author} {\bibfnamefont {A.}~\bibnamefont
  {Dragan}}, \bibinfo {author} {\bibfnamefont {J.}~\bibnamefont {Doukas}},
  \bibinfo {author} {\bibfnamefont {E.}~\bibnamefont {Mart\'in-Mart\'inez}}, \
  and\ \bibinfo {author} {\bibfnamefont {D.~E.}\ \bibnamefont {Bruschi}},\
  }\href@noop {} {\bibfield  {journal} {\bibinfo  {journal} {Class. Quantum
  Grav.}\ }\textbf {\bibinfo {volume} {30}},\ \bibinfo {pages} {235006}
  (\bibinfo {year} {2013})}\BibitemShut {NoStop}%
\bibitem [{\citenamefont {Doukas}\ \emph {et~al.}(2013)\citenamefont {Doukas},
  \citenamefont {Brown}, \citenamefont {Dragan},\ and\ \citenamefont
  {Mann}}]{Doukas13}%
  \BibitemOpen
  \bibfield  {author} {\bibinfo {author} {\bibfnamefont {J.}~\bibnamefont
  {Doukas}}, \bibinfo {author} {\bibfnamefont {E.~G.}\ \bibnamefont {Brown}},
  \bibinfo {author} {\bibfnamefont {A.}~\bibnamefont {Dragan}}, \ and\ \bibinfo
  {author} {\bibfnamefont {R.~B.}\ \bibnamefont {Mann}},\ }\href {\doibase
  10.1103/PhysRevA.87.012306} {\bibfield  {journal} {\bibinfo  {journal} {Phys.
  Rev. A}\ }\textbf {\bibinfo {volume} {87}},\ \bibinfo {pages} {012306}
  (\bibinfo {year} {2013})}\BibitemShut {NoStop}%
\bibitem [{\citenamefont {Berman}\ \emph {et~al.}(1993)\citenamefont {Berman},
  \citenamefont {with~contributions from}, \citenamefont {Barton},
  \citenamefont {Carmichael}, \citenamefont {Childs}, \citenamefont
  {Gabrielse}, \citenamefont {Tan}, \citenamefont {Haroche}, \citenamefont
  {Raimond}, \citenamefont {Hinds}, \citenamefont {Kimble}, \citenamefont
  {Meystre}, \citenamefont {Wilkens}, \citenamefont {Mossberg}, \citenamefont
  {Lewenstein},\ and\ \citenamefont {et~al}}]{CavityQED}%
  \BibitemOpen
  \bibfield  {author} {\bibinfo {author} {\bibfnamefont {P.~R.}\ \bibnamefont
  {Berman}}, \bibinfo {author} {\bibnamefont {with~contributions from}},
  \bibinfo {author} {\bibfnamefont {G.}~\bibnamefont {Barton}}, \bibinfo
  {author} {\bibfnamefont {H.}~\bibnamefont {Carmichael}}, \bibinfo {author}
  {\bibfnamefont {J.}~\bibnamefont {Childs}}, \bibinfo {author} {\bibfnamefont
  {G.}~\bibnamefont {Gabrielse}}, \bibinfo {author} {\bibfnamefont
  {J.}~\bibnamefont {Tan}}, \bibinfo {author} {\bibfnamefont {S.}~\bibnamefont
  {Haroche}}, \bibinfo {author} {\bibfnamefont {J.}~\bibnamefont {Raimond}},
  \bibinfo {author} {\bibfnamefont {E.}~\bibnamefont {Hinds}}, \bibinfo
  {author} {\bibfnamefont {H.}~\bibnamefont {Kimble}}, \bibinfo {author}
  {\bibfnamefont {P.}~\bibnamefont {Meystre}}, \bibinfo {author} {\bibfnamefont
  {M.}~\bibnamefont {Wilkens}}, \bibinfo {author} {\bibfnamefont
  {T.}~\bibnamefont {Mossberg}}, \bibinfo {author} {\bibfnamefont
  {M.}~\bibnamefont {Lewenstein}}, \ and\ \bibinfo {author} {\bibfnamefont
  {G.~R.}\ \bibnamefont {et~al}},\ }\href
  {http://www.amazon.com/Quantum-Electrodynamics-Advances-Molecular-Optical/dp/0120922452%3FSubscriptionId%3D0JYN1NVW651KCA56C102%26tag%3Dtechkie-20%26linkCode%3Dxm2%26camp%3D2025%26creative%3D165953%26creativeASIN%3D0120922452}
  {\emph {\bibinfo {title} {Cavity Quantum Electrodynamics (Advances in Atomic,
  Molecular and Optical Physics)}}}\ (\bibinfo  {publisher} {Academic Press},\
  \bibinfo {year} {1993})\BibitemShut {NoStop}%
\bibitem [{\citenamefont {Blais}\ \emph {et~al.}(2004)\citenamefont {Blais},
  \citenamefont {Huang}, \citenamefont {Wallraff}, \citenamefont {Girvin},\
  and\ \citenamefont {Schoelkopf}}]{Blais04}%
  \BibitemOpen
  \bibfield  {author} {\bibinfo {author} {\bibfnamefont {A.}~\bibnamefont
  {Blais}}, \bibinfo {author} {\bibfnamefont {R.-S.}\ \bibnamefont {Huang}},
  \bibinfo {author} {\bibfnamefont {A.}~\bibnamefont {Wallraff}}, \bibinfo
  {author} {\bibfnamefont {S.~M.}\ \bibnamefont {Girvin}}, \ and\ \bibinfo
  {author} {\bibfnamefont {R.~J.}\ \bibnamefont {Schoelkopf}},\ }\href@noop {}
  {\bibfield  {journal} {\bibinfo  {journal} {Phys. Rev. A}\ }\textbf {\bibinfo
  {volume} {69}},\ \bibinfo {pages} {062320} (\bibinfo {year}
  {2004})}\BibitemShut {NoStop}%
\bibitem [{\citenamefont {You}\ and\ \citenamefont {Nori}(2011)}]{You11}%
  \BibitemOpen
  \bibfield  {author} {\bibinfo {author} {\bibfnamefont {J.~Q.}\ \bibnamefont
  {You}}\ and\ \bibinfo {author} {\bibfnamefont {F.}~\bibnamefont {Nori}},\
  }\href@noop {} {\bibfield  {journal} {\bibinfo  {journal} {Nature}\ }\textbf
  {\bibinfo {volume} {474}},\ \bibinfo {pages} {589} (\bibinfo {year}
  {2011})}\BibitemShut {NoStop}%
\bibitem [{\citenamefont {Wallraff}\ \emph {et~al.}(2004)\citenamefont
  {Wallraff}, \citenamefont {Schuster}, \citenamefont {Blais}, \citenamefont
  {Frunzio}, \citenamefont {Huang}, \citenamefont {Majer}, \citenamefont
  {Kumar}, \citenamefont {Girvin},\ and\ \citenamefont
  {Schoelkopf}}]{meandering1}%
  \BibitemOpen
  \bibfield  {author} {\bibinfo {author} {\bibfnamefont {A.}~\bibnamefont
  {Wallraff}}, \bibinfo {author} {\bibfnamefont {D.~I.}\ \bibnamefont
  {Schuster}}, \bibinfo {author} {\bibfnamefont {A.}~\bibnamefont {Blais}},
  \bibinfo {author} {\bibfnamefont {L.}~\bibnamefont {Frunzio}}, \bibinfo
  {author} {\bibfnamefont {R.~S.}\ \bibnamefont {Huang}}, \bibinfo {author}
  {\bibfnamefont {J.}~\bibnamefont {Majer}}, \bibinfo {author} {\bibfnamefont
  {S.}~\bibnamefont {Kumar}}, \bibinfo {author} {\bibfnamefont {S.~M.}\
  \bibnamefont {Girvin}}, \ and\ \bibinfo {author} {\bibfnamefont {R.~J.}\
  \bibnamefont {Schoelkopf}},\ }\href {http://dx.doi.org/10.1038/nature02851}
  {\bibfield  {journal} {\bibinfo  {journal} {Nature}\ }\textbf {\bibinfo
  {volume} {431}},\ \bibinfo {pages} {162} (\bibinfo {year}
  {2004})}\BibitemShut {NoStop}%
\bibitem [{\citenamefont {van Loo}\ \emph {et~al.}(2013)\citenamefont {van
  Loo}, \citenamefont {Fedorov}, \citenamefont {Lalumière}, \citenamefont
  {Sanders}, \citenamefont {Blais},\ and\ \citenamefont
  {Wallraff}}]{meandering2}%
  \BibitemOpen
  \bibfield  {author} {\bibinfo {author} {\bibfnamefont {A.~F.}\ \bibnamefont
  {van Loo}}, \bibinfo {author} {\bibfnamefont {A.}~\bibnamefont {Fedorov}},
  \bibinfo {author} {\bibfnamefont {K.}~\bibnamefont {Lalumière}}, \bibinfo
  {author} {\bibfnamefont {B.~C.}\ \bibnamefont {Sanders}}, \bibinfo {author}
  {\bibfnamefont {A.}~\bibnamefont {Blais}}, \ and\ \bibinfo {author}
  {\bibfnamefont {A.}~\bibnamefont {Wallraff}},\ }\href {\doibase
  10.1126/science.1244324} {\bibfield  {journal} {\bibinfo  {journal}
  {Science}\ }\textbf {\bibinfo {volume} {342}},\ \bibinfo {pages} {1494}
  (\bibinfo {year} {2013})},\ \Eprint
  {http://arxiv.org/abs/http://www.sciencemag.org/content/342/6165/1494.full.pdf}
  {http://www.sciencemag.org/content/342/6165/1494.full.pdf} \BibitemShut
  {NoStop}%
\bibitem [{\citenamefont {Hoi}\ \emph {et~al.}(2011)\citenamefont {Hoi},
  \citenamefont {Wilson}, \citenamefont {Johansson}, \citenamefont {Palomaki},
  \citenamefont {Peropadre},\ and\ \citenamefont {Delsing}}]{Hoi11}%
  \BibitemOpen
  \bibfield  {author} {\bibinfo {author} {\bibfnamefont {I.-C.}\ \bibnamefont
  {Hoi}}, \bibinfo {author} {\bibfnamefont {C.~M.}\ \bibnamefont {Wilson}},
  \bibinfo {author} {\bibfnamefont {G.}~\bibnamefont {Johansson}}, \bibinfo
  {author} {\bibfnamefont {T.}~\bibnamefont {Palomaki}}, \bibinfo {author}
  {\bibfnamefont {B.}~\bibnamefont {Peropadre}}, \ and\ \bibinfo {author}
  {\bibfnamefont {P.}~\bibnamefont {Delsing}},\ }\href {\doibase
  10.1103/PhysRevLett.107.073601} {\bibfield  {journal} {\bibinfo  {journal}
  {Phys. Rev. Lett.}\ }\textbf {\bibinfo {volume} {107}},\ \bibinfo {pages}
  {073601} (\bibinfo {year} {2011})}\BibitemShut {NoStop}%
\bibitem [{\citenamefont {Hoi}\ \emph {et~al.}(2012)\citenamefont {Hoi},
  \citenamefont {Palomaki}, \citenamefont {Lindkvist}, \citenamefont
  {Johansson}, \citenamefont {Delsing},\ and\ \citenamefont {Wilson}}]{Hoi12}%
  \BibitemOpen
  \bibfield  {author} {\bibinfo {author} {\bibfnamefont {I.-C.}\ \bibnamefont
  {Hoi}}, \bibinfo {author} {\bibfnamefont {T.}~\bibnamefont {Palomaki}},
  \bibinfo {author} {\bibfnamefont {J.}~\bibnamefont {Lindkvist}}, \bibinfo
  {author} {\bibfnamefont {G.}~\bibnamefont {Johansson}}, \bibinfo {author}
  {\bibfnamefont {P.}~\bibnamefont {Delsing}}, \ and\ \bibinfo {author}
  {\bibfnamefont {C.~M.}\ \bibnamefont {Wilson}},\ }\href {\doibase
  10.1103/PhysRevLett.108.263601} {\bibfield  {journal} {\bibinfo  {journal}
  {Phys. Rev. Lett.}\ }\textbf {\bibinfo {volume} {108}},\ \bibinfo {pages}
  {263601} (\bibinfo {year} {2012})}\BibitemShut {NoStop}%
\bibitem [{\citenamefont {Delsing}()}]{PerDelsing}%
  \BibitemOpen
  \bibfield  {author} {\bibinfo {author} {\bibfnamefont {P.}~\bibnamefont
  {Delsing}},\ }\href@noop {} {\bibinfo  {journal} {in a private
  communication.}\ }\BibitemShut {NoStop}%
\bibitem [{\citenamefont {Zupanzic}(2013)}]{Zupancic}%
  \BibitemOpen
\bibfield  {journal} {  }\bibfield  {author} {\bibinfo {author} {\bibfnamefont
  {P.}~\bibnamefont {Zupanzic}},\ }\href@noop {} {\emph {\bibinfo {title}
  {Dynamic Holography and Beamshaping using Digital Micromirror Devices}}}\
  (\bibinfo  {publisher} {LMU M\"unich, Grainer Lab Harvard},\ \bibinfo {year}
  {2013})\BibitemShut {NoStop}%
\bibitem [{\citenamefont {Simon}(2000)}]{Simon00}%
  \BibitemOpen
  \bibfield  {author} {\bibinfo {author} {\bibfnamefont {R.}~\bibnamefont
  {Simon}},\ }\href@noop {} {\bibfield  {journal} {\bibinfo  {journal} {Phys.
  Rev. Lett}\ }\textbf {\bibinfo {volume} {84}},\ \bibinfo {pages} {2726}
  (\bibinfo {year} {2000})}\BibitemShut {NoStop}%
\end{thebibliography}%

\end{document}